\documentclass[a4paper,12pt]{article}
\usepackage[english]{babel}

\usepackage[utf8x]{inputenc}
\usepackage[T1]{fontenc}
\usepackage[letterpaper, margin=1in]{geometry} 
\usepackage{multirow}
\usepackage{graphicx}
\usepackage{amsmath}
\usepackage{bbm}
\usepackage{indentfirst}  
\usepackage{makecell}
\usepackage{threeparttable}
\usepackage{setspace}
\usepackage{adjustbox}
\usepackage{array}
\usepackage{subfigure}
\usepackage{caption}
\usepackage{color}
\usepackage[table,xcdraw]{xcolor}
\usepackage{appendix}
\usepackage{natbib}
\usepackage{authblk}
\usepackage{amssymb,amsmath,bm,mathrsfs,makeidx,amsfonts,graphicx,amsthm,multirow}
\usepackage[figuresright]{rotating}
\usepackage{longtable}
\usepackage{algorithm}
\usepackage{algorithmic}
\usepackage{booktabs}
\newcolumntype{?}{!{\vrule width 1pt}}
\numberwithin{equation}{section}

\usepackage[pdftex,colorlinks=true,hypertexnames=false]{hyperref}
\definecolor{darkblue}{rgb}{0,0,.6}
\hypersetup{citecolor=darkblue,linkcolor=darkblue,urlcolor=darkblue}

\DeclareMathOperator{\sgn}{sgn}

\DeclareMathOperator{\tr}{tr}

\newcommand{\beqs}{\begin{eqnarray}}
\newcommand{\eeqs}{\end{eqnarray}}
\newcommand{\beqsn}{\begin{eqnarray*}}
	\newcommand{\eeqsn}{\end{eqnarray*}}

\newtheorem{remark}{Remark}

\newtheorem{theorem}{Theorem}

\newtheorem{lemma}{Lemma}
\newtheorem{assumption}{Assumption}

\newcommand{\probP}{\text{I\kern-0.15em P}}

\newcommand{\bbA}{{\boldsymbol A}}

\newcommand{\bbB}{{\boldsymbol B}}

\newcommand{\bbR}{{\boldsymbol R}}

\newcommand{\bbV}{{\boldsymbol V}}

\begin{document}
\onehalfspacing
\begin{sloppypar}
\title{Adaptive Multi-task Learning for Multi-sector Portfolio Optimization
}

\author[a]{Qingliang Fan}
\author[b,c]{Ruike Wu\thanks{\noindent Corresponding to: School of Economics, Shanghai University of Finance and Economics, Shanghai, China.
Email: ruikewu@foxmail.com (R. Wu). }}
\author[d]{Yanrong Yang}

\affil[a]{Department of Economics, The Chinese University of Hong Kong}
\affil[b]{School of Economics, Shanghai University of Finance and Economics}
\affil[c]{Key Laboratory of Mathematical Economics (SUFE), Ministry of Education}
\affil[d]{College of Business and Economics, The Australian National University}

\date{}

\maketitle
\vspace{-2.5em}
\begin{abstract}
Accurately capturing the transfer of information across multiple sectors to enhance model estimation is both significant and challenging in multi-sector portfolio optimization involving a large number of assets in different sectors. Within the framework of factor modeling, we propose a novel data-adaptive multi-task learning methodology that quantifies and learns the relatedness among the principal temporal subspaces (spanned by factors) across multiple sectors. This approach improves the estimation of multiple factor models and thus enhances multi-sector portfolio optimization. A novel and easy-to-implement algorithm, termed projection-penalized principal component analysis, is developed to accomplish the multi-task learning procedure. We establish asymptotic properties for both the estimators from the multi-task factor model and the associated multi-task portfolio risk and Sharpe ratio estimators. Our simulation study, and empirical applications based on the data of the Russell 3000 index components with the Fama-French industrial sectors, demonstrate the advantages of the newly proposed multi-task learning methodology.  
\end{abstract}

\noindent {\bf{Keywords}}: Multi-task Learning, Adaptive learning, Minimum variance portfolio, Factor model, Projection subspace

\newpage
\section{Introduction}

Modern financial markets are vast and complex. The equity market is conventionally classified by multiple industry sectors, such as manufacturing, financial services, and technology. There are also diverse asset classes, including equities (stocks), fixed income (bonds), commodities (gold, oil) and cash equivalents (treasury bills, money market funds). Investors nowadays consider interconnected markets that enable cross-border trading and investment, with convenient access through electronic platforms to international markets such as the US, Europe, and Asia.  When constructing investment portfolios, funds typically take into account various sectors and markets in order to achieve diversification and risk reduction. Most fund products are associated with a certain “sector,” which can be broadly interpreted as a country-specific financial market, an asset class, an industrial sector or subsector, or an investment theme such as ESG. Fund managers are required to conduct extensive research and analysis on firms within a given sector or theme. The task involves determining portfolio weights, implementing rebalancing strategies, and diversifying across assets within the sector, with the objective of achieving favorable Sharpe ratio performance and effective risk control. It is likely that some sectors share similar underlying driving forces of asset returns. In this paper, we are interested in the following question: \emph{can we utilize the potential but unknown homogeneities across different sectors to improve the estimation of factor models and portfolio performance?}

Homogeneous information contained in the common factors across different sectors has already been documented in the financial literature.  \cite{griffin2002fama}  investigates whether the well-known Fama-French three factors, introduced by \cite{fama1993common}, are global or country-specific in nature. Its empirical findings suggest that incorporating foreign factors into a domestic factor model yields only a marginal increase in explanatory power, indicating that country-specific factors across regions share a substantial amount of common information.  \cite{kose2003international} employ a Bayesian dynamic latent factor model to estimate common components in macroeconomic aggregates across 60 countries spanning seven major regions. Their findings reveal that a common global factor is an important source of volatility in most countries, providing evidence for the existence of a global business cycle and indicating the presence of substantial homogeneous information across countries. \cite{bollerslev2014stock} investigate how the variance risk premium predicts stock market returns in the U.S. and several other countries through country-specific regressions. They uncover strong predictability and nearly identical patterns across countries, indicating the presence of homogeneous information across multiple country-specific markets.  Additionally, it is natural to classify individual stocks based on their broad industry categories. However, it is worth noting that many commonly used factors in stock markets are constructed without accounting for industry distinctions (for example, \cite{fama1993common,fama2015five}), indicating that different industrial sectors may share substantial information.

Portfolio allocation across multiple related asset groups naturally gives rise to a multi-task econometric problem. The flexible quantification and adaptive estimation of shared information across tasks can lead to improved performance.  On one hand, pooling data across all sectors can improve estimation efficiency when all sectors share exactly the same factors, but may introduce model misspecification when heterogeneity is present.  In the financial literature, assets worldwide can be grouped into different regions, with returns in each region driven by region-specific factors.  Within a given market, assets can be categorized into sectors, and returns in each sector may be influenced by sector-specific factors \citep[and references therein]{hawawini2003performance}. These indicate the presence of heterogeneous information across sectors. On the other hand, estimating each sector independently is robust to heterogeneity but may lead to efficiency loss, as it fails to fully utilize the shared information across sectors, especially in data-rich environments. Therefore, it is crucial to develop a data-adaptive method that effectively leverages information across sectors while avoiding the model misspecification inherent in pooling, thereby balancing these two extremes.


In this study, we consider a multi-task learning framework that simultaneously analyzes multiple financial sectors by leveraging potentially homogeneous information embedded across them, with the goal of enhancing performance in multi-sector portfolio allocation. Specifically, we propose a projection-penalized factor model that jointly performs factor analysis across multiple sectors by exploiting potential homogeneity in the common factors through penalizing the distance between the projection matrices spanned by these factors. The degree of penalization depends on a regularization parameter, which is selected via data-driven cross-validation, rendering the proposed multi-task procedure data-adaptive and allowing it to automatically adjust to the degree of homogeneity across sectors. To solve the proposed multi-task optimization problem, an algorithm that is easy to implement and computationally efficient is developed. Based on the covariance matrix estimators for each sector, we then construct sector-specific minimum variance portfolios, which have been widely studied in high-dimensional portfolio allocation \citep{demiguel2009generalized, ledoit2017nonlinear, ding2021high, fan2022time},  for investment purposes.
Under suitable regularization conditions, we establish the asymptotic properties of the proposed estimators for factor loadings, common factors, the error covariance matrix, and the inverse return covariance matrix in each sector,  and further establish the consistency of the  portfolio risk estimator and Sharpe ratio estimator. When the regularization parameter shrinks to zero faster than a certain rate (see details in Section \ref{sec: theorem}), the convergence rates of the proposed estimators reduce to those of classical methods \citep{bai2003, fan2013large}, which rely only on within-sector data.
Moreover, under complete homogeneity, where all sectors share the same common factors, the proposed multi-task estimators achieve the same or even faster rates as those obtained by classical estimation based on fully pooled data.\footnote{When all sectors are of comparable size, the multi-task method and the pooled-data-based method achieve the same convergence rate. However, when sector sizes are heterogeneous---in particular, when a few sectors contain substantially more assets than others---the multi-task method attains a faster rate for the sectors having less assets. See more details in Lemma \ref{lemma 2 homo}.} 
Simulation results show that the proposed portfolio strategy outperforms both the individual strategy (which analyzes each sector independently) and the pooling strategy (which is based on fully pooled data) in terms of overall \footnote{The overall performance is measured as the average across all considered sectors.} estimation error, out-of-sample risk error, and out-of-sample Sharpe ratio error when the considered sectors share a proportion of information (partial homogeneity). The performance of the proposed estimators and portfolios is comparable to, or even better than, that of the individual and pooling strategies under both complete heterogeneity and homogeneity, demonstrating the adaptiveness of the proposed strategy.
Finally, we conduct empirical analysis based on individual stocks in the Russell 3000 index and the well-known Fama-French industry classification. The empirical results show that the proposed multi-task portfolio strategy achieves the best overall performance in terms of out-of-sample Sharpe ratio and risk, outperforming the individual strategy and two pooling strategies, the latter of which are based on fully pooled data. Moreover, for investors focusing on a specific sector, the empirical results show that the proposed multi-task strategy can deliver improved performance by incorporating information from other highly related sectors.

The contributions of this paper are summarized as follows: (1) We propose a novel projection-penalized factor model that jointly performs factor analysis for multiple groups, utilizing the potential commonality in their latent factors, and is useful in various applications including finance, economics, psychometrics, and environmental science. This extends the POET estimator in \cite{fan2013large} to a multi-sector setting. 
We quantify the ``interconnections'' among different sectors by the distance between the  subspace spanned by common factors across multiple sectors.  (2) We develop a homogeneity-guided iterative principal component algorithm to solve the proposed multi-task optimization problem, which is easy to implement in practice and computationally efficient.
(3)  The proposed method is data-adaptive to the degree of homogeneity, which is characterized by a regularization coefficient selected via cross-validation.   (4) We establish the theoretical properties of the multi-task learning-based estimators, including those for factor loadings, common factors, the error covariance matrix, the inverse return covariance matrix, portfolio risk, and the Sharpe ratio. Moreover, we show that under complete homogeneity, the proposed multi-task estimators achieve the same or even faster rates as those obtained by classical estimation based on fully pooled data. 
(5) To our knowledge, a multi-sector version of minimum variance portfolio is novel in the literature, which can achieve better overall performance across all considered sectors.  In particular, by incorporating sectors that are highly related to a specific sector of interest, the multi-task learning-based portfolio can achieve better performance than that of the individual strategy. 

This paper is related to the multi-task analysis.
Common multi-task approaches regularize task-specific parameters through regularization terms \citep{evgeniou2005learning}, as well as cluster or pooled datasets based on their similarities \citep{crameer2008, jacob2008clustered, mcdonald2016new}.
Recently, 
\cite{duan2023adaptive} and \cite{knight2023multi} study the multi-task learning for linear regression problem  by using $l_2$-norm-based penalties;   \cite{huang2025optimal} and \cite{xu2025multitask} study multi-task learning and  bandits under sparse heterogeneity, where the source/task-associated parameters are equal to a global parameter plus a sparse task-specific term. Earlier studies on multi-task learning can be found in the references of the aforementioned works. To the best of our knowledge, no existing research has investigated multi-task learning for  portfolio decision analysis in high-dimensional setups.

This paper also connects to the studies of transfer learning and integrative analysis. 
Recently, \cite{tang2016fused} introduce a fused lasso regression method  to identify heterogeneity patterns of coefficients and to merge the homogeneous parameter clusters across multiple datasets; 
\cite{fang2018integrative} propose an integrative sparse principal component analysis approach for multiple independent datasets, where the similarities between datasets are measured by the distance between factor loadings or by the signs of the loadings;  
\cite{li2022transfer} consider estimation and prediction of a high-dimensional linear regression in the setting of transfer
learning;
\cite{tian2023transfer} study the transfer learning problem under high-dimensional generalized linear models;
\cite{li2023transfer} focus on high-dimensional Gaussian graphical models;
\cite{cai2024transfer} employ minimax rate theory to study the estimation of the mean  of random functions;
\cite{gu2024robust} propose an angle-based approach to exploit the similarity between two linear regression models; \cite{cao2023transfer} study the application of transfer learning techniques  to  financial portfolio optimization problem by penalizing the distance between the pre-trained portfolio and the transferred portfolio; \cite{morstedt2024cross} propose a novel covariance estimator and the corresponding portfolio strategy based on non-linear shrinkage, where historically optimal shrinkage parameters are transferred to the current estimation task. Different from them, we focus on the analysis of multi-task factor model and adaptively utilize the potential interconnected features across different sectors for portfolio allocation.

This paper is closely associated with high-dimensional portfolio allocation. The Markowitz mean-variance portfolio is elegant in theory, but has limitations in high-dimensional models due to estimation errors in the covariance matrix and mean returns \citep{michaud1989markowitz, jagannathan2003risk, demiguel2009generalized, ao2019approaching}. Therefore, many works have been devoted to developing high-dimensional portfolio allocation methods. For example, in recent studies, \cite{ao2019approaching} propose the MAXSER mean-variance portfolio by reformulating the Markowitz problem as a regression problem and imposing an $\ell_1$ penalization; \cite{li2022synthetic} further extend the MAXSER framework (without factor investing) by incorporating the Fama-French factor model and using  synthetic regression. \cite{chen2025cross} and \cite{ao2025robust} further extend MAXSER to allow for the dimension to exceed the sample size and to accommodate heteroskedasticity and heavy tails. \cite{ding2021high} propose a unified framework for high-dimensional minimum-variance portfolios under statistical factor models. In particular, when the minimum risk diverges rapidly, their strategy reduces to the minimum-variance portfolio based on the POET estimator \citep{fan2013large}. \cite{fan2022time} consider time-varying factor loadings and propose time-varying minimum-variance portfolios. To address the infeasibility of distributionally robust portfolios in high dimensions \citep{blanchet2022distributionally}, \cite{wu2024uncertainty} extend their framework under latent factor models and propose a high-dimensional version of distributionally robust mean-variance portfolios. Based on random matrix theory, \cite{bodnar2018estimation} and \cite{bodnar2022optimal} develop portfolio combining rule which are available in high dimensional setup. \cite{lassance2024combination} investigate how to best combine the sample mean–variance portfolio with the naive equally weighted portfolio to
optimize out-of-sample performance in high dimensional situation. We note that all the high-dimensional portfolio allocation mentioned above are restricted to a single sector and do not focus on utilizing cross-sector information.


This paper is organized as follows. Section \ref{sec: estimation} introduces the multi-sector factor model and the associated estimation procedure. Section \ref{sec: theorem} presents the main theoretical results. Simulation studies are reported in Section \ref{sec: simulation}. Section \ref{sec: empirical} conducts empirical analyses based on the components of the Russell 3000 index and the Fama-French industrial classification. Section \ref{sec: con} concludes the paper. The detailed steps for solving the multi-task optimization problem \eqref{PCA loss Multi}, along with the full proofs of the main theoretical results, are provided in the Supplementary Material.

In this paper,  $\bbA^\top$ denotes the transpose of $\bbA$; 
$\|\bbA \|$, $\|\bbA\|_1$, and $\|\bbA\|_F$ denote the spectral norm, $\ell_1$ norm, and Frobenius norm of matrix $\bbA$, which are defined as $\theta_1(\bbA)$, $\sum_{ij}|
\bbA_{ij}|$, and $\sqrt{\tr(\bbA^\top \bbA)}$, respectively, where $\theta_i(\bbA)$ is the $i$-th largest eigenvalue of a matrix $\bbA$, and $\tr(\bbA)$ is the trace of $\bbA$. If $\bbA$ is a vector, both  $\|\bbA\|$ and $\|\bbA\|_F$ equal the Euclidean norm.

\section{Multi-sector Factor Model}\label{sec: estimation}
\subsection{MVP Preliminaries}
In this paper, we consider an adaptive homogeneity learning for multi-sectors based on minimum variance portfolio  (MVP) in high-dimensional setup with the help of factor structure. The minimum variance portfolio optimization has a surging appearance in recent studies on portfolio management \citep{jagannathan2003risk, demiguel2009generalized, fan2012vast, bodnar2018estimation, ding2021high, fan2022time} pioneered by Markowitz's mean-variance model \citep{markowitz1952portfolio}. Specifically, the classic MVP problem is:
\begin{equation}
\label{eq gmvp}
\begin{split} \min \ w^{\top}\Sigma_r w,\ \ \ \ \ \ \\ \text{s.t.} \ w^{\top} 1_p = 1. \ \ \ \ \ 
\end{split}
\
\end{equation}
where $\Sigma_r$ is a $p$-dimensional population covariance matrix of asset returns, $w$ is a $p\times1$ vector of asset weights in the portfolio, and $1_p$ is a $p\times1$ vector of 1's. 
	The analytical solution for MVP weight is:
	\begin{equation}
	\label{analytic solution for gmvp}
	w^* = \frac{\Sigma_r^{-1} 1_p}{1_p^{\top}\Sigma_r^{-1} 1_p}.
	\end{equation}
	The variance of MVP and the corresponding Sharpe ratio (SR) are:
	\begin{equation}
	\label{analytic solution for minimum variance}
	R_{\min}= {w^{*}}^{\top} \Sigma_r w^* = \frac{1}{1_p^{\top}\Sigma_r^{-1} 1_p},
	\end{equation}
	\begin{equation}
	\label{analytic solution for SR}
	\text{SR} = \frac{{w^{\ast}}^{\top}\mu_r}{ \sqrt{ {w^{*}}^{\top} \Sigma_r w^*}}=\frac{1_p^{\top}\Sigma_r^{-1} \mu_r }{\sqrt{1_p^{\top} \Sigma_r^{-1} 1_p}},
	\end{equation}
	respectively, where $\mu_r$ is a vector of expected excess returns (over a risk-free rate) of $p$ stocks.

\subsection{Two-sector Case}
We first consider the simplest multi-sector model with two sectors. Throughout this section, the terms ``sector'' and ``market'' are used interchangeably to refer to the conglomeration of data from one sector. We collect two datasets $R^{(1)}$ and $R^{(2)}$ from two different markets (denoted market 1 and market 2), the underlying structures of two markets can be arbitrarily related.  We are interested to see whether the common information shared between the two markets can lead to improved statistical estimation for markets 1 and 2, on average, compared to using only the individual datasets $R^{(1)}$ and $R^{(2)}$. 
	
In asset pricing studies, the factor model is the main workhorse \citep{fama1992crosssection,bai2002determining,fan2013large} that draws from the capital asset pricing model (CAPM). The theoretical work of arbitrage pricing theory (APT, \citealp{ross1976arbitrage}) laid the foundation for approximate factor models, which conveniently decompose the return data matrix to a low-rank component and a sparse component \citep{Chamberlain83}.
Specifically, we model the (excess) assets return from two markets $m = 1,2$ by   
\begin{eqnarray}\label{yr(1)}
r_{it}^{(m)}=b_{i}^{(m)\top}F_t^{(m)}+e_{it}^{(m)}, \ i=1, 2, \ldots, p_m; \ \ t=1, 2, \ldots, T_m; \ \ m = 1,2, 
\end{eqnarray}
where $r_{it}^{(m)}$ is the (excess of the risk-free) return of asset $i$ in market $m$ at time $t$, $F_t^{(m)}$ is the $K_m$-dimensional vector of common factors in market $m$ at time $t$, $K_m$ is the number of common factors in market $m$, $b_{i}^{(m)} = (b_{i1}^{(m)},\ldots,b_{iK_m}^{(m)})^\top$ is a $K_m$ dimensional vector of factor loading, which captures the relationship between common factors and asset $i$ in market $m$; and $e_{it}^{(m)}$ is the idiosyncratic error.  In practice, factors can be either observed or not, and in this paper, we assume unobserved factors as the default setting. 
	
Let $r_t^{(m)}=\left(r_{1t}^{(m)}, \ldots, r_{p_m t}^{(m)}\right)^{\top}$, $B^{(m)}=\left(b_{1}^{(m)}, \ldots, b_{p_m}^{(m)}\right)^{\top}$, and $e_t=\left(e_{1t}, \ldots, e_{p_m t}\right)^{\top}$. The vector form of model (\ref{yr(1)}) is given by 
\begin{eqnarray}\label{vecform}
r_t^{(m)}=B^{(m)}F_t^{(m)}+e_t^{(m)}, \ \ \ t=1, 2, \ldots, T_m, \ \ m = 1,2. 
\end{eqnarray}

Based on the factor structure, the information in asset returns is captured by three components: the underlying common factors, the factor loadings, and the idiosyncratic errors. It is well known that factor loadings and idiosyncratic errors are specific to individual assets, while the common factors drive the co-movement of a basket of assets, effectively reflecting market- and sector-level information. 
Motivated by the principle of the factor model, it is natural to use a ``distance measure'' between the common factors of two markets to quantify their similarity in information, and hence to assess the usefulness of joint modeling for estimation. 
In this paper, we use the distance measured by the Frobenius norm between the projection matrices spanned by the common factors of two markets to quantify the degrees of homogeneity between the two markets.  
Using the distance between projection matrices has two main merits: (1) it accommodates all cases of differences in factor spaces, including the case with the same number of factors but different factor structures, as well as cases with varying numbers of common factors across markets; (2) the factor space is invariant to rotations of the common factors, which is particularly useful when the common factors are unobservable and need to be estimated.

 As a result, we consider the following projection-penalized optimization problem, which  utilizes the similarity information between two markets, to estimate the factor loading and common factors in each sector: 
\begin{align}
&\min_{B^{(1)},F^{(1)},B^{(2)},F^{(2)}}  L(B^{(1)},B^{(2)},F^{(1)},F^{(2)})  \nonumber
\\ & := \frac{1}{T}\sum^{T}_{t=1} \left\| r_t^{(1)} -  {B^{(1)}}{F}_t^{(1)}\right\|^2 + \frac{1}{T}\sum^{T}_{t=1} \left\| r_t^{(2)} -  {B^{(2)}}{F}_t^{(2)}\right\|^2 + \frac{\lambda}{T} \left\| P^{(1)} - P^{(2)} \right\|_F^2, 
\label{PCA loss}
\end{align}
subject to the identification conditions that $F^{(m)\top}F^{(m)}/T = I_K$ and $B^{(m)\top}B^{(m)}$ is a diagonal matrix for  $m = 1,2$, where $F^{(m)} = (F_1^{(m)},\cdots,F_{T_m}^{(m)})^\top$. Here,  $P^{(m)}$ is the projection matrix of common factors,  defined as   $P^{(m)} = F^{(m)}(F^{(m)\top}F^{(m)})^{-1}F^{(m)\top}$, $m = 1,2$. We note that projection-based penalization requires the data from the two markets to share the same sample size $T$  since we penalize the discrepancy between $T^{(m)}\times T^{(m)}$ projection matrices.  This restriction is mild and easily satisfied, as two datasets can be viewed as partitions of a single pooled market dataset, categorized by sectors such as industry classifications or by the markets in which the equities are traded. As a benefit of requiring the same sample size, the numbers of assets in the two markets are allowed to differ, which facilitates practical implementation, since matching the number of stocks across multiple sectors can sometimes be challenging. Lastly, under the identification condition, the projection matrix can be simplified to $F^{(m)}F^{(m)\top}/T$, which helps simplify the theoretical derivation.

The positive constant $\lambda$ is called the regularization parameter, which controls the trade-off between the prediction error we make on the two factor models and the complexity of the solution. Intuitively, when $\lambda = 0$, the optimal solutions are the same as those derived from classic principal component analysis (PCA) of each market independently, and $\lambda$  makes two tasks unrelated. When $\lambda$ is sufficiently large,  $\hat{F}^{(1)}$ and  $\hat{F}^{(2)}$ are forced to be close to each other, effectively enforcing the two market models to be identical.
 Note that, $\hat{B}^{(1)}$ is not expected to be close to $\hat{B}^{(2)}$, as factor loadings are asset-specific and can differ significantly across the two sectors. 

\begin{remark}
The distance between two projection matrices provides a natural measure of relatedness between the factor structures of two markets. Suppose that market 1 and market 2 have $K_1$ and $K_2$ common factors, where $K_1$ may differ from $K_2$, and the common factors in the two markets may also differ. For an asset $i$ in market 1, the term $P^{(1)}(r_{i1},\ldots,r_{iT})^\top$ represents the component of returns explained by the common factors in market 1, while
 $P^{(2)}(r_{i1},\ldots,r_{iT})^\top$ represents the component explained by the factor space of market 2.  If the projection matrices of the two markets are similar, these systematic components will also be close, implying that the factor structures of the two markets capture similar patterns of systematic variation in asset returns. This indicates that the two markets share substantial common information in terms of risk pricing. 
\end{remark}

 \begin{remark}
    As an alternative to the Frobenius norm based penalty, we can penalize the difference between two projection matrices using the $\ell_1$-norm. This approach may yield solutions for which the corresponding projection matrices are identical in expectation. However, the main advantage of the Frobenius norm based penalty is that it leads to an explicit solution to the optimization problem \eqref{PCA loss} (see Algorithm \ref{algonew-mul}), which is often preferred by financial practitioners and is computationally simpler. In addition, since the goal is to exploit (potentially) homogeneous information to improve estimation rather than to enforce exact equality, analysis based on the Frobenius norm based penalty suffices.
    \label{remark l1 penalty}
 \end{remark}
 
\subsection{Multi-sector Case}
Now consider the multi-sector case. We assume that there are $M$ sectors,  and the (excess) assets return from $M$ sectors can all be modeled by factor model shown in (\ref{yr(1)}).  The sector number $M$ is fixed. The goal is to simultaneously estimate the factor models for each sector while accounting for the relatedness among various markets. Accordingly, for each sector $m \in [1,2,\ldots,M]$, return data are collected for $p_m$ assets over a sample size $T$, and thus we consider the following multi-task minimization problem:
\begin{align}
&\min_{B^{(1)},F^{(1)},\ldots,B^{(M)},F^{(M)}}  L(B^{(1)},F^{(1)},\ldots,B^{(M)},F^{(M)})  \nonumber
\\ &= \frac{1}{T}\sum_{m=1}^{M}\sum^{T}_{t=1} \left\| r_t^{(m)} -  {B^{(m)}}{F}_t^{(m)}\right\|^2 + \frac{\lambda}{T}\sum_{m=1}^{M}\sum_{m^\prime > m} \left\| P^{(m)} - P^{(m^\prime)}\right\|_F^2,
\label{PCA loss Multi}
\end{align}
subject to the identification conditions that $F^{(m)\top}F^{(m)}/T = I_K$ and $B^{(m)\top}B^{(m)}$ is a diagonal matrix for  $m = 1,\ldots,M$.

First, in multi-task optimization \eqref{PCA loss Multi}, one regularization parameter $\lambda$ is imposed on all projection discrepancies, as no prior information indicates which markets are similar. We point out that some techniques, such as the cross-validation method applied in this paper for tuning parameter selection, allow the data to reveal relationships among the markets under consideration, thereby avoiding subjective judgments by practitioners. Additionally, from a statistical inference perspective, only a single tuning parameter, $\lambda$, needs to be selected, which is appealing. It is often unclear whether and how a task can be better addressed by incorporating information from other tasks, and the proposed strategy offers a novel way to tackle this challenge.

\begin{remark}
In the multi-task optimization problem \eqref{PCA loss Multi}, we assign equal weights to the prediction errors across sectors for simplicity.   In practice, given differences across tasks such as the number of assets within each sector, a weighted formulation of the problem is also feasible. 
\end{remark}

The optimization problem \eqref{PCA loss Multi} can be solved using the first-order arguments, as detailed in our online supplementary material. Based on the analytical solution, the following Algorithm \ref{algonew-mul} is proposed to   numerically estimate factor loadings and common factors for multiple sectors iteratively: 

\begin{algorithm}[H]
\caption{Estimation procedure for the multi-task optimization problem \eqref{PCA loss Multi}.}
\label{algonew-mul}
\begin{algorithmic}[1]

\STATE Set $i \leftarrow 1$.
\STATE Initialize $B^{(m)}$ and $F^{(m)}$ for $m=1,\ldots,M$. 
\COMMENT{Default: $F^{(m),0}=\sqrt{T}\,\text{eig}_K(R^{(m)\top}R^{(m)})$, 
$B^{(m),0}=T^{-1}R^{(m)}F^{(m),0}$, where $\text{eig}_K(\cdot)$ denotes the eigenvectors corresponding to the $K$ largest eigenvalues.}

\REPEAT

\FOR{$m=1,\ldots,M$}

\STATE Compute
\[
V^{(m),i} = \frac{1}{T}R^{(m)\top}R^{(m)} 
+ \frac{2\lambda}{T}\sum_{j<m}P^{(j),i}
+  \frac{2\lambda}{T}\sum_{j>m}P^{(j),i-1} - \frac{2\lambda}{T}(M-1)I_T,
\]
where $P^{(m),i-1}=T^{-1}F^{(m),i-1}F^{(m),i-1\top}$, and $I_T$ is the $T$ dimensional identity matrix.

\STATE Update $
F^{(m),i}=\sqrt{T}\text{eig}_K(V^{(m),i})$ and
$B^{(m),i}=T^{-1}R^{(m)}F^{(m),i}.$

\ENDFOR

\STATE $\Delta = |L^{(i)}-L^{(i-1)}|$, where $
L^{(i)} = L(B^{(1),i},F^{(1),i},\ldots,B^{(M),i},F^{(M),i})$.

\STATE $i \leftarrow i+1$.

\UNTIL{$i>MaxI$ or $\Delta < tol$}

\end{algorithmic}
\end{algorithm}

Algorithm \ref{algonew-mul} requires five inputs: the return data of multiple sectors ($R^{(m)}, m =1,\ldots,M$); the number of common factors $K_m$ for each sector, the tuning parameter $\lambda$, the maximum steps $MaxI$, and the tolerance level $tol$. It is worth mentioning that even though the proposed penalty involves pairwise comparisons across all sectors, leading to a computational complexity that scales quadratically with the number of sectors ($O(M^2)$), in practice the proposed Algorithm \ref{algonew-mul} is computationally efficient and typically converges within a few iterations.
	
\subsection{Covariance Matrix Estimation}
After obtaining estimations $\widehat{B}^{(m)}$ and $\widehat{F}_t^{(m)}$, we can further compute the estimated residuals $\hat{e}_t^{(m)} = r_t  - \widehat{B}^{(m)}\widehat{F}_t^{(m)}$ for each sector. For estimating the error covariance matrix $\Sigma_{e}^{(m)}$ of each market,  we follow \cite{fan2013large}, \cite{WANG202153}, and \cite{wu2024uncertainty}, and impose approximate sparsity assumption on $\Sigma_{e}^{(m)} = (\sigma_{e,ij}^{(m)})_{p\times p}$, that is, for some $q \in [0,1)$,
	\begin{align}
	\label{sparse kp}
	k_{q,m}=\max_{i\leq p_m}\sum_{j\leq p_m}|\sigma_{e,ij}^{(m)}|^{q}.
	\end{align}
	does not grow too fast as $p_m\rightarrow \infty$. In particular, $k_{q,m}$ is the maximum number of non-zero elements in each row when $q = 0$. Hence, following the work of \cite{cai2011adaptive}, we can apply the adaptive thresholding estimation to the off-diagonal elements in sample covariance matrix, and the sparse error covariance matrix estimator is given by
	\begin{align}
	\hat{\Sigma}_e^{(m)} = (\hat{\sigma}_{\hat{e},ij}^{(m)})_{p\times p}, \quad \hat{\sigma}_{\hat{e},ij}^{(m)} = \left\{ \begin{array}{ll} s_{\hat{e},ii}^{(m)} & i=j \\ \xi_{ij}(s_{\hat{e},ij}^{(m)}) & i\neq j \end{array} \right.
	\end{align}
	where $s_{\hat{e},ij}^{(m)}= T^{-1}\sum_{s=1}^T\hat{e}_{is}^{(m)}\hat{e}_{js}^{(m)}$ is the $(i,j)$-th element of sample covariance matrix of $\hat{e}_t^{(m)}$, $\xi_{m,ij}(\cdot)$ is a shrinkage function\footnote{The general $\xi_{m,ij}(\cdot)$ includes many commonly-used thresholding functions such as soft thresholding ($\xi_{m,ij}(z) = \sgn(z)(|z|-\tau_{ij}^{(m)})^{+}$, $(z)^+ = max\{z,0\}$) and hard thresholding ($\xi_{m,ij}(z) = z\mathbb{I}(|z|\geq \tau_{ij}^{(m)})$). } satisfying
	$\xi_{m,ij}(z)=0$ if $|z| \leq \tau_{ij}^{(m)}$, and $|\xi_{m,ij}(z)-z|\leq \tau_{ij}^{(m)}$, where $\tau_{ij}^{(m)}$ is an entry-adaptive positive threshold, defined as  $\tau_{ij}^{(m)} = C_\tau \vartheta_{T,m}\sqrt{\hat{\psi}_{ij}^{(m)}}$, where $C_\tau > 0$ is a sufficiently large constant, $\vartheta_{T,m} = \sqrt{\log p_m/T}$, 
	and $\hat{\psi}_{ij}^{(m)}= T^{-1}\sum_{t=1}^T\left( \hat{e}_{it}^{(m)}\hat{e}_{jt}^{(m)}-{s}_{\hat{e},ij}^{(m)}\right)^2$.

With the estimated sparse error covariance matrix  $\hat{\Sigma}_e^{(m)}$ in hand, we can immediately obtain the estimated covariance matrix of asset return for each sector  $m = 1,\ldots, M$ as follows:
\begin{align}
  \hat{\Sigma}_r^{(m)} = \hat{B}^{(m)}\hat{\Sigma}_f^{(m)}\hat{B}^{(m)\top} +  \hat{\Sigma}_e^{(m)},  
  \label{our covariance estimation}
\end{align}
where $\hat{\Sigma}_f^{(m)}$ is sample covariance matrix of estimated factors $\widehat{F}_t^{(m)}$.

Finally, let $\hat{\Omega}_r$ denote the inverse of $\hat{\Sigma}_r$, then the estimated minimum variance portfolio in each market is further given by 
\begin{align}
    \hat{w}^{(m)} = \frac{\hat{\Omega}_r^{(m)} 1_{p_m}}{1_{p_m}^{\top}\hat{\Omega}_r^{(m)} 1_{p_m}},
    \label{eq estimated portfolio}
\end{align}
with the estimated portfolio risk $\hat{R}_{min} = 1/(1_{p_m}^\top \hat{\Omega}_r^{(m)} 1_{p_m})$ and the estimated Sharpe ratio $1_{p_m}^\top \hat{\Omega}_r^{(m)} \hat{\mu}_r^{(m)}/\sqrt{1_{p_m}^\top \hat{\Omega}_r^{(m)} 1_{p_m}}$, where $\hat{\mu}_r^{(m)} = \hat{B}^{(m)}\sum_t \hat{F}_t^{(m)}/T$.

For factor-based methodologies, we also need to determine the number of common factors in each market. In this paper, we simply apply the standard and commonly used approach proposed by \cite{bai2002determining} to determine the number of common factors $K_m$. In details,  for each sector, we estimates the number of common factors by 
\begin{align}
\hat{K}_m = \arg \min_{0 \leq K \leq K_{\max} } \log\left(\frac{1}{p_mT}\left\| R^{(m)} - T^{-1}R^{(m)}\hat{F}^{(m)}(K)\hat{F}^{(m)\top}(K)  \right\|^2_F \right) + Kg(T,p_m) 
\label{eq: factor number esti}
\end{align}
where $K_{\max}$ is a prescribed upper bound for the number of factors, and the $i^{th}$ column of matrix $\hat{F}^{(m)}(K)/\sqrt{T}$  corresponds to the $i^{th}$ largest eigenvalues of matrix $R^{(m)\top} R^{(m)}$, and $g(T,p_m)$ is a penalty function of $p_m$ and $T$. The assumptions imposed in this paper are sufficient to guarantee the consistency of the estimated number of factors.  Following the study of  \cite{fan2013large}, two suggested penalty functions 
$$
g_1(T,p_m) = \frac{(p_m + T)}{p_m T}\log\left(\frac{p_m T}{p_m + T}\right), \quad g_2(T,p_m) = \frac{(p_m + T)}{p_m T}\log(\min(p_m,T))
$$ are selected.
Let $\hat{K}_{m,1}$ and $\hat{K}_{m,2}$ denote the estimated value by applying the above two penalty functions, respectively, and we set the final $\hat{K}_m$ as the integer part of $({\hat{K}_{m,1}+\hat{K}_{m,2}})/2$.

\subsection{Selection of Regularization Parameter}
\label{sec: choice of regu}

 In practice, we use $k$-fold cross-validation to select value for regularization parameter $\lambda$. Specifically, for return data $R^{(m)}$ in each market $ m = 1,2,\ldots, M$, we randomly split $R^{(m)}$ into $k$ groups to form $k$ validation sets, denoted by $\{R_{1}^{(m)},\ldots,R_{k}^{(m)}\}$. For each validation set $R_{j}^{(m)}, j = 1,\ldots,k$, the remaining observations in $R^{(m)}$ are used as the training set. For each training set and a given value of $\lambda$, we obtain estimates of the factor loadings and common factors for all sectors by solving the optimization problem \eqref{PCA loss Multi};  Based on these estimates, the minimum-variance portfolios for each sector are constructed and denoted by $\hat{w}_i^{(m)}(\lambda)$, $i=1,\ldots,k$. Then, we select $\lambda_{cv}$  to minimize the out-of-sample portfolio risk on the validation set as follows:
\begin{equation}
    \lambda_{cv} = \arg \min_{\lambda}  \frac{1}{kM}\sum_{m=1}^M\sum_{i=1}^{k} Risk(\hat{w}_i^{(m)}(\lambda),R_{i}^{(m)})
    \label{cv for lambda}
\end{equation}
where the out-of-sample risk $Risk(\hat{w}_i^{(m)},R_{i}^{(m)}) = (T_i -1)^{-1}\sum_{t=1}^{T_i}\left(\hat{w}_i^{(m)\top}r_{i,t}^{(m)} - \bar{r}^{(m)}_i \right)^2$, and $\bar{r}^{(m)}_i(\lambda) = T_i^{-1}\sum_{t=1}^{T_i}\hat{w}_i^{(m)\top}r_{i,t}^{(m)}$, $T_i$ is the number of  observations in $i^{th}$ validation set, and $r_{i,t}^{(m)}$ is the return data in $i^{th}$ validation set. A similar selection procedure can be found in \cite{ding2021high}. Note that the criterion (\ref{cv for lambda}) minimizes the average out-of-sample portfolio risk across all considered sectors. Thus, this tuning parameter selection is adopted from a holistic perspective encompassing all sectors. If investors are interested only in one specific sector, it is natural to select a $ \lambda$ that minimizes the out-of-sample risk solely for that sector.

In empirical applications, specifying an appropriate value for $C_\tau$ is required to carry out adaptive thresholding estimation for error covariance matrices. In this paper, we apply a data-driven $N$-fold cross-validation procedure to determine values for $C_\tau$ in each sector. Similar procedures can be found in \cite{bickel2008covariance}, \cite{cai2011adaptive}, and \cite{fan2013large}.
More precisely, with the residual $\hat{e}_t^{(m)}$ from our multi-task estimation procedure, we randomly divide it into two subsets, which are denoted as $\{\hat{e}_t^{(m)}\}_{t\in A} $ and $\{\hat{e}_t^{(m)}\}_{t\in B} $. Subset $A$ is used for training and subset $B$ is used for validation purpose. Therefore, we choose the threshold $C_\tau$ by minimizing the following objective function over a compact interval:
\begin{align}
C_\tau^* = \arg \min_{C_{\tau,min}<C_\tau \leq C_{\tau, max}} \frac{1}{N}\sum_{j=1}^{N} \left\|\hat{\Sigma}_e^{(m),A,j}(C_\tau) - S_e^{(m),B,j} \right\|^2_F
    \label{eq:Ctau select}
	\end{align}
	where $C_{\tau,min}$ is the minimum constant that guarantees the positive definiteness of $\hat{\Sigma}_e^{(m),A,j}(C_\tau)$, $C_{\tau, max}$ is a large constant such that $\hat{\Sigma}_e^{(m),A,j}(C_\tau)$ is diagonal matrix, $\hat{\Sigma}_e^{(m),A,j}(C_\tau)$ is the sparse thresholding residual covariance estimator by using subset $A$ in $j^{th}$ loop with threshold $C_\tau$ for sector $m$, and $S_e^{(m),B,j} $ is sample covariance matrix by using subset $B$ in $j^{th}$ loop for sector $m$. In our numerical analysis, we take $T(A)= [2T/3]$. 

\section{Asymptotic Theory}\label{sec: theorem}
In this section, we establish the asymptotic properties for a series of developed estimators for factors, factor loadings, the covariance matrix, the minimum risk  and the Sharpe ratio.

\begin{assumption}
    All the eigenvalues of the $K_m \times K_m$ matrix $p_m^{-1}B^{(m)\top}B^{(m)}$ are bounded away from both 0 and $\infty$ as $p_m \rightarrow \infty$ for $m = 1,\ldots, M$.
    \label{assum loading}
\end{assumption}
\begin{assumption}
    For $m = 1, 2,\ldots, M$,
 (a) $\{e_t^{(m)}, F_t^{(m)} \}_{t\geq 1}$ are all strictly stationary, and  $\text{E}\left(e_{it}^{(m)}\right) = \text{E}\left(e_{it}^{(m)}F_{jt}^{(m)}\right) = 0$ for $i\leq p_m, j\leq K_m$ and $t\leq T$. (b) There are constants $C_1, C_2 > 0$ such that $\lambda_{\min}\left(\Sigma_e^{(m)}\right) > C_1$, $\left\| \Sigma_e^{(m)} \right\|_1 < C_2$ and $\min_{i\leq p_m,j\leq p_m} \text{var}(e_{it}^{(m)}e_{jt}^{(m)}) > C_1$. (c) $\text{E}\left(F_t^{(m)}F_t^{(m)\top}\right) = I_{K_m}$. (d)  There are $\gamma_1^{(m)}, \gamma_2^{(m)} > 0$ and $\varsigma_1,\varsigma_2 > 0$ such that for any $s>0$, $i\leq p_m$ and $j \leq K_m$, $\text{Pr}\left(\left\vert e_{it}^{(m)}\right\vert>s\right) \leq exp\{-(s/\varsigma_1)^{\gamma_1^{(m)}} \}$ and  $\text{Pr}\left(\left\vert F_{it}^{(m)}\right\vert >s\right) \leq exp\{-(s/\varsigma_2)^{\gamma_2^{(m)}} \}$. 
\label{assum relation}
\end{assumption}

\begin{assumption}
For $m = 1,\ldots, M$, there exists $\gamma_3^{(m)} > 0$ such that $3[\gamma_1^{(m)}]^{-1} + 1.5[\gamma_2^{(m)}]^{-1} + [\gamma_3^{(m)}]^{-1} > 1$, and $C > 0$ satisfying, for all $T_1 \in \mathbb{Z}^+$, $\alpha(T_1) \leq \text{exp}(-CT_1^{\gamma_3^{(m)}})$, where $\alpha(T_1)$ is the $\alpha$-mixing coefficient defined on $\sigma$-algebra generated by $\{(F_t^{(m)},e_t^{(m)}): t\leq 0 \}$ and $\{(F_t^{(m)},e_t^{(m)}): t \geq T_1 \}$.
    \label{assum mixing}
\end{assumption}

\begin{assumption}
 For $m = 1,\ldots, M$, there exists $C > 0$ such that, for $i \leq p_m$, $t\leq T$ and $s \leq T$, (a) $\left\| b_i^{(m)} \right\| < C$, (b) $\text{E}\left(p_m^{-1/2} \left(e_s^{(m)\top} e_t^{(m)} - \text{E}(e_s^{(m)\top} e_t^{(m)}) \right) \right)^4 < C$, (c) $\text{E}\left\| p_m^{-1/2}\sum_{i=1}^p b_i^{(m)} e_{it}^{(m)} \right\| < C$.
 \label{assum regularization}
\end{assumption}

Assumption \ref{assum loading} is common in the literature of approximate factor model, it requires  the common factors to be pervasive, that is, to impact a non-vanishing
proportion of individual return series.  It indicates that the first $K_m$ eigenvalues of $\Sigma_r^{(m)}$  grow at a  rate  of $p_m$.  
Assumption \ref{assum relation}(a) assumes that  both $e_t^{(m)}$ and $F_t^{(m)}$ are strictly stationary, and are uncorrelated.
Assumption \ref{assum relation}(b), combined with Assumption \ref{assum loading}, establishes a spiked structure for the covariance matrix of returns. The constraint on $\ell_1$ norm of error covariance matrix is employed to ensure a consistent estimation of the number of factors \citep{bai2002determining,bai2003, fan2013large}.  Assumption \ref{assum relation}(c) normalizes the second moment of the common factors to the identity matrix, which is commonly adopted in the latent factor model literature; see, for example, \cite{fan2013large}, \cite{li2022integrative}, \cite{wu2024uncertainty}. Since we focus on covariance-based portfolio allocation, which is invariant to rotations of the factors, the results are not essentially affected.
The exponential tail condition in Assumption \ref{assum relation}(d) and the strong mixing condition in Assumption \ref{assum mixing} enable the application of large deviation theory to some terms such as $T^{-1}\sum_{t=1}^T F_t^{(m)}e_{it}^{(m)}$ and $T^{-1}\sum_{t=1}^T e_{it}^{(m)} e_{jt}^{(m)}$, see  \cite{fan2011high} for more details.
Assumption \ref{assum regularization}  regulates some moment conditions, which are  needed to estimate consistently the transformed common factors as
well as the factor loadings, also see \cite{bai2002determining}, \cite{bai2009panel}, and \cite{fan2013large}.

To better analyze the statistical properties of estimation from  multi-task optimization \eqref{PCA loss Multi},  we further define $$\varphi_i = \sum_{j\neq i}^M \left\| P^{(i)} - P^{(j)} \right\|,$$
which measures the population relatedness of sector $i$ with respect to other sectors.
The relatedness of sector $i$ within the group of considered sectors is measured by the sum of the distances between the projection matrix spanned by common factors of sector $i$ and the those of other sectors in the group. It is evident that a smaller $\varphi_i$ indicates stronger relatedness with other sectors, and therefore reflects more homogeneous information among groups. When $\varphi_i = 0$, the common factors across all sectors share  the same information.\footnote{Factors in different sectors may differ, but they span the same factor space.} Note that $\varphi_i = O(M)$ since the distance between projection matrix is measured by spectral norm. Further let $\hat{\Theta}_T^{(m)}$ denote the $K_m \times K_m$ diagonal matrix of the first $K_m$ largest eigenvalues of matrix 
$$
 \frac{1}{T}R^{(m)\top}R^{(m)}  +  \frac{2\lambda}{T} \sum_{m^\prime \neq m}^{M} \frac{1}{T}F^{(m)}F^{(m)\top} - \frac{2\lambda}{T}(M-1)I_{T},
$$
for $m = 1,\ldots,M$, and define $H^{(m)} = T^{-1}\hat{\Theta}_{T}^{(m)-1}F^{(m)\top}\hat{F}^{(m)}B^{(m)\top} B^{(m)} $. The following Lemma \ref{lemma 1} provides the convergence results for common factors and factor loadings.
\begin{lemma}
 Let $(\gamma^{(m)})^{-1} = 3(\gamma_1^{(m)})^{-1} + 1.5(\gamma_2^{(m)})^{-1}+(\gamma_3^{(m)})^{-1} + 1$. Suppose that for $m =1,\ldots, M$, $\log p_m = o(T^{\gamma^{(m)}/6})$, $T = o(p_m^2)$, $\lambda/(p_m T^{1/4}) = o(1)$, $\lambda >0$, and Assumptions \ref{assum loading}-\ref{assum regularization} hold.
Then we have
 \begin{itemize}
 \item [(a)] $\max_{t\leq T} \left\| \hat{F}_t^{(m)} - H^{(m)} F_t^{(m)}  \right\| = O_p\left(\frac{T^{1/4}}{\sqrt{p_m}} + \frac{1}{\sqrt{T}} + \iota_{m}\right).$
     \item[(b)] $\max_{i\leq p_m}\left\| \hat{b}_i^{(m)} -H^{(m)} b_i^{(m)}\right\| = O_p\left(\sqrt{\frac{\log p_m}{T}}  + \frac{1}{\sqrt{p_m}} + \iota_{m}\right).$
 \end{itemize}
 where $\iota_m =\lambda T^{-1/4}p_m^{-1}(M + \varphi_m)$
 \label{lemma 1}
\end{lemma}

Lemma \ref{lemma 1} presents the asymptotic properties of estimators for common factors and factor loadings in each sector. 
 It is clear that both common factor and factor loading estimations are consistent with the corresponding population counterparts.
Compared to the classic asymptotic results shown in \cite{fan2013large}, the additional term $\iota_m$, which is related to the number of sectors $M$ and the relatedness of sector $m$ to other sectors $\varphi_m$, arises from accounting for multiple datasets. 
 In particular, when $\lambda = 0$, the optimization problem for each sector is independent of the others and the multi-task analysis reduces to a separate analysis for each sector; In this case, the convergence rates coincide with those shown in Theorem 4 of \cite{fan2013large}.  Moreover, when $\lambda = o(\sqrt{p_m T})$ and $\lambda = o(T^{-1/4}p_m \log p_m)$, the additional term $\iota_m$ is asymptotically dominated by the other terms, so that the penalty  does not affect the asymptotic results. However, when either of these rate conditions fails to hold, the penalty term actually affect the final convergence results.

Let $\omega_{T,m} = \sqrt{\log p_m/T} + p_m^{-1/2} + \lambda T^{-1/4}p_m^{-1}(M+\varphi_m)$, the following Theorem \ref{theorem error cov} shows the convergence rate of the sparse  error covariance matrix (and its inverse) estimator:
\begin{theorem}
 Under the conditions of Lemma \ref{lemma 1}, $\omega_{T,m}^{1-q}\kappa_{q,m} = o(1)$. Then, for a sufficiently large constant $C_\tau>0$ in the adaptive threshold $\tau_{ij}^{(m)} = C_\tau\vartheta_{T,m}\sqrt{\hat{\psi}_{ij}^{(m)}}$, for $m = 1,\ldots, M$, we have
\begin{align*}
\left\| \hat{\Sigma}_e^{(m)} - \Sigma_e^{(m)} \right\| = O_p(\omega_{T,m}^{1-q}\kappa_{q,m}),
\quad
\left\| \hat{\Omega}_e^{(m)} - \Omega_e^{(m)} \right\| = O_p(\omega_{T,m}^{1-q}\kappa_{q,m}),    
\end{align*}
where $\Omega_e^{(m)}$ and $\hat{\Omega}_e^{(m)}$ are the inverse of $\Sigma_e^{(m)}$ and $\hat{\Sigma}_e^{(m)}$, respectively. 
\label{theorem error cov}
\end{theorem}

Next, we give the rate of convergence for the inverse of covariance matrix (precision matrix) estimation in each sector.

\begin{theorem}
    Under the conditions of Theorem \ref{theorem error cov}, for $m = 1,\ldots, M$, we have 
   $$
   \left\|  \hat{\Omega}_r^{(m)}  - \Omega_r^{(m)} \right\|  = O_p(\omega_{T,m}^{1-q}\kappa_{q,m}), 
   $$
where $\hat{\Omega}_r^{(m)}$ and $\Omega_r^{(m)}$ are the inverse of $\hat{\Sigma}_r^{(m)}$  and $\Sigma_r^{(m)}$, respectively.
\label{theorem sigma r}
\end{theorem}

 Theorem \ref{theorem sigma r} establishes the convergence rate of the inverse of the estimated covariance matrix to its population counterpart in each sector under the spectral norm, which plays a key role in deriving the convergence results for portfolio risk and Sharpe ratio estimations. Combine with Theorem \ref{theorem error cov}, 
it is clear that the convergence rate of the inverse of return covariance estimation is
the same as that of the error covariance matrix estimation, which  indicates that the estimation
accuracy of $\hat{\Sigma}_e^{(m)}$ plays an important role in  minimum variance portfolio allocation.

Next, we study the asymptotic properties of the portfolio risk estimator and the Sharpe ratio estimator. To facilitate the analysis, we impose the following assumption.
\begin{assumption}\label{assu: risk} For $m= 1,\ldots, M$, the minimum risk $R_{\min}^{(m)}=\frac{1}{1_p^\top\Omega_{r}^{(m)}1_p}\asymp p_m^{1-\eta}$, where $\eta$ is a  constant satisfying $p_m^{2-\eta}\omega_{T,m}^{1-q}\kappa_{q,m}=o(1)$. 
	\end{assumption}
	
\begin{assumption}\label{assu: sr}  For $m= 1,\ldots, M$,  suppose the term $1_p^\top\Omega_{r}^{(m)}\mu^{(m)} \asymp p_m^{1-\phi}$, where $\phi$ is a  constant satisfying $p_m^{\phi}\omega_{T,m}^{1-q}\kappa_{q,m} =o(1)$.
  \end{assumption}

Assumptions \ref{assu: risk} and \ref{assu: sr} require the minimum risk $R_{min}$ and  ${\mathbf{1}_p^\top \Sigma_r^{-1} \mu}$ are of the order of powers of $p$. Similar conditions can be found, e.g., in \cite{fan2015risks}, \cite{ding2021high}, \cite{caner2023sharpe}, and \cite{fan2022time}. A simple example illustrating our assumptions can be found in subsection 2.4.1 of  \cite{ding2021high}.

\begin{theorem}
    \label{theorem risk convergence}
    Under conditions of Theorem \ref{theorem sigma r}, and suppose Assumption \ref{assu: risk} holds. For $m= 1,\ldots, M$, we  have the convergence rate for minimum risk estimator:
    $$
\left|\frac{\hat{R}_{min}^{(m)}}{R_{min}^{(m)} } -1 \right| = O_p\left(p_m^{2-\eta} \omega_{T,m}^{1-q}\kappa_{q,m} \right) = o_p(1).    
    $$
\end{theorem}

In the special case of $\eta = 2$, which is often assumed in the high-dimensional portfolio literature (see, e.g., \cite{fan2015risks}, \cite{caner2023sharpe}, and \cite{caner2025portfolio}), the convergence rate reduces to that of the error covariance matrix estimation. 
With a similar ratio criterion, we further assess the Sharpe ratio estimator. 
\begin{theorem}
\label{theorem SR convergence}
Under conditions of Theorem \ref{theorem sigma r}, and suppose Assumptions \ref{assu: risk} and \ref{assu: sr} hold. For $m = 1,\ldots, M$, we have the convergence rate for Sharpe ratio estimator:
    $$
\left|\frac{\widehat{SR}^{(m)}}{SR^{(m)}} -1 \right| = O_p\left((p_m^{2-\eta}+p_m^\phi)\omega_{T,m}^{1-q}\kappa_{q,m} \right) = o_p(1).    
    $$
\end{theorem}
Theorem \ref{theorem SR convergence} demonstrate the consistency of Sharpe ratio estimator. Similarly, in the special case of $\phi = 0$ (see, e.g., \cite{callot2021nodewise} and \cite{caner2023sharpe}), together with $\eta = 2$, the convergence rate also reduces to that of the error covariance matrix estimation.

Next, we consider a special case of complete homogeneity in which the assets in all markets share the same factors, that is $F^{(1)} =\ldots = F^{(M)} =: F$. We will show that under complete homogeneity, the estimators for each sector obtained from the proposed multi-task learning approach achieve the same or even faster convergence rate than those obtained by pooling all data together. Note that, since all sectors share the same factors, it follows that $K_m =: K$, $\gamma_2^{(m)}=:\gamma_2$, and $\gamma_3^{(m)}=:\gamma_3$ for $m=1,\ldots, M$.
\begin{lemma}
Under the complete homogeneous case where  $F = F^{(1)}=\ldots = F^{(M)}$, and suppose that $\gamma_1^{(m)} = \gamma_1$ for $m=1,\ldots,M$. Define  $\gamma^{-1} = 3\gamma_1^{-1} + 1.5\gamma_2^{-1}+\gamma_3^{-1} + 1$. Under Assumptions 1-4, $\log \left(\sum_{i=1}^M p_m\right) = o(T^{\gamma/6})$, $T = o\left((\sum_{i=1}^Mp_i)^2\right)$, 
$\lambda > 0$ and $p_mT/\lambda = o(1)$.  Then we have
 \begin{itemize}
 \item [(a)] $\max_{t\leq T} \|\hat{F}_t^{(m)} - H^{all}F_t\| = O_p\left(\frac{1}{\sqrt{T}}+\frac{T^{1/4}}{\sqrt{\sum_{i=1}^M p_i}}\right),$
 \item [(b)]
$\max_{i\leq p_m} \left\|\hat{b}_i^{(m)} - H^{all}b_i^{(m)} \right\| = O_p\left(\sqrt{\frac{\log p_m}{T}} + \frac{1}{\sqrt{\sum_{i=1}^M p_i}}\right)$,
 \end{itemize}
where $H^{all} = (1/T)\bbV^{-1}\hat{F}^{(m)\top}F\bbB^\top \bbB$, $\bbB = (B^{(1)\top}, B^{(2)\top},\ldots,B^{(M)\top})^\top$, and $\bbV^{-1}$ denotes the
diagonal matrix of the first $K$ largest eigenvalues of  $\bbR^{all}=\sum_{m=1}^M\frac{1}{T}R^{(m)\top}R^{(m)}$.
\label{lemma 2 homo}
\end{lemma}

Lemma \ref{lemma 2 homo} establishes the asymptotic properties of the proposed multi-task estimator under complete homogeneity. Under this setting, a natural alternative strategy for investors is to pool the return data from all sectors together and apply the classical PCA to the resulting $\left(\sum_i p_i\right) \times T $ data matrix to estimate the common factors and factor loadings. For this alternative method, by using the Theorem 4 of \cite{fan2013large}, the pooled PCA estimators  $\hat{F}^{pool}_t$, $\hat{b}_i^{pool}$ satisfy $
\max_{t} \|\hat{F}_t^{pool} - H^{pool}F_t\| = O_p\left(T^{-1/2}+T^{1/4}\left(\sum_{i=1}^M p_i\right)^{-1/2}
\right)$ and $\max_{i} \left\|\hat{b}_i^{pool} - H^{pool}b_i^{pool} \right\| = O_p\left(T^{-1/2}\sqrt{\log \left(\sum_i p_i\right)} + \left(\sum_{i=1}^M p_i\right)^{-1/2}\right)$,  where $b_i^{pool}$ is the population factor loading of asset $i$, and  $H^{pool}$ is some rotation matrix.
It is evident that, under complete homogeneity, the proposed factor estimator achieves the same convergence rate as the pooled estimator. Furthermore, for factor loadings, the two methods differ in the logarithmic term: the rate of multi-task estimator involves $\log p_m$, whereas the that of pooled estimator involves $\log(\sum_mp_m)$. The two rates coincide when all sectors are of comparable size, since
two terms differ only by $\log M$ which can be negligible. However, when sector sizes are heterogeneous, particularly when a few sectors are much larger than others, the multi-task estimator yields a faster rate for the smaller sectors.
We note that the derivation of Lemma \ref{lemma 2 homo} requires $p_m T/\lambda = o(1)$,  which does not satisfy the conditions assumed in Lemma \ref{lemma 1}. Hence, Lemma \ref{lemma 2 homo} is not a special case of Lemma \ref{lemma 1} when $\varphi_m = 0$ for $m=1,\ldots, M$.
Developing a unified result that depends on the population relatedness, including the complete homogeneity case, is challenging, and we leave this for future work.  



Similarly, corresponding to Theorems \ref{theorem error cov}, \ref{theorem sigma r}, \ref{theorem risk convergence} and \ref{theorem SR convergence}, the following theorem establishes the convergence rate of error covariance matrix, inverse of return covariance matrix, Sharpe risk and Sharpe ratio estimators.

\begin{theorem}
Under the conditions of Lemma \ref{lemma 2 homo}, $\omega_{ho,T,m}^{1-q}\kappa_{q,1} = o(1)$ with $\omega_{ho,T,m} = \sqrt{\log p_m/T} +(\sum_i p_i)^{-1/2}$. Then, for a sufficiently large constant $C_\tau>0$ in the threshold $\tau_{ij}^{(m)} = C_\tau\vartheta_{T,m}\sqrt{\hat{\psi}_{ij}^{(m)}}$, for $m = 1,\ldots, M$, we have   \begin{align*}
\left\| \hat{\Sigma}_e^{(m)} - \Sigma_e^{(m)} \right\| = O_p(\omega_{ho,T,m}^{1-q}\kappa_{q,m}), 
\   
\left\| \hat{\Omega}_r^{(m)} - \Omega_r^{(m)} \right\| = O_p(\omega_{ho,T,m}^{1-q}\kappa_{q,m}).
\end{align*} 
\label{theorem homo cov}
\end{theorem}

\begin{theorem}
\label{theorem SR convergence homo}
Under conditions of Theorem \ref{theorem homo cov}, and suppose Assumptions \ref{assu: risk} and \ref{assu: sr} hold. For $m = 1,\ldots, M$, we have the convergence rate for Sharpe risk and Sharpe ratio estimators:
\begin{align*}
\left|\frac{\hat{R}_{min}^{(m)}}{R_{min}^{(m)} } -1 \right| = O_p\left(p_m^{2-\eta} \omega_{ho,T,m}^{1-q}\kappa_{q,m} \right) = o_p(1), 
\  \left|\frac{\widehat{SR}^{(m)}}{SR^{(m)}} -1 \right| = O_p\left((p_m^{2-\eta}+p_m^\phi)\omega_{ho,T,m}^{1-q}\kappa_{q,m} \right) = o_p(1).
\end{align*}
\end{theorem} 


\section{Simulation Study}\label{sec: simulation}
In this section, we evaluate the newly proposed method via Monte Carlo Simulation.
	
\subsection{Model Setup}

\begin{table}[htbp!]
\caption{Simulation results for the scenario in which all sectors have the same number of assets (50) and the sample size is 250. EcovE is the error covariance estimation error, INVcovE is the inverse of return covariance matrix estimation error, WE is weight estimation error, RiskE is risk estimation error, SRE is the Sharpe ratio estimation error, OOS RiskE is the distance between out-of-sample estimated portfolio risk and that of the population portfolio, OOS SRE is the distance between out-of-sample estimated portfolio Sharpe ratio and that of the population portfolio. Replace 0-6 denote the number of replacement steps.
“Replace 0” corresponds to complete homogeneity, where all sectors share the same two basic factors. At each step, a block of five factors is replaced by newly generated ones. “Replace 6” corresponds to maximal heterogeneity, where all sector-specific factors are fully distinct.}
\label{Table Simu1}
\centering
\tabcolsep 0.056in
\begin{tabular}{cccccccc}
    \toprule Measures      & Individual & Multi-task & Pooled-sector &  & Individual & Multi-task & Pooled-sector \\ \midrule
          & \multicolumn{3}{c}{Replace 0}           &  & \multicolumn{3}{c}{Replace 1}           \\
EcovE     & 5.00E-04   & 2.99E-04   & 3.83E-04      &  & 4.99E-04   & 3.23E-04   & 4.88E-03      \\
INVcovE   & 7650       & 4791       & 6476          &  & 7650       & 5383       & 12435         \\
WE        & 0.5223     & 0.4599     & 0.4876        &  & 0.5214     & 0.4803     & 0.7190        \\
RiskE     & 1.60E-04   & 1.21E-04   & 1.45E-04      &  & 1.60E-04   & 1.25E-04   & 2.55E-04      \\
SRE       & 0.0167     & 0.0155     & 0.0161        &  & 0.0404     & 0.0395     & 0.0586        \\
OOS RiskE & 2.52E-04   & 1.73E-04   & 2.10E-04      &  & 2.51E-04   & 1.88E-04   & 5.55E-04      \\
OOS SRE   & 0.0334     & 0.0311     & 0.0318        &  & 0.0346     & 0.0321     & 0.0717        \\ \midrule
          & \multicolumn{3}{c}{Replace 2}           &  & \multicolumn{3}{c}{Replace 3}           \\
EcovE     & 5.06E-04   & 3.31E-04   & 8.85E-03      &  & 4.94E-04   & 3.24E-04   & 1.33E-02      \\
INVcovE   & 7649       & 5581       & 18659         &  & 7638       & 5608       & 19174         \\
WE        & 0.5186     & 0.4803     & 0.9307        &  & 0.5191     & 0.4790     & 1.1214        \\
RiskE     & 1.59E-04   & 1.26E-04   & 3.79E-04      &  & 1.59E-04   & 1.29E-04   & 9.12E-04      \\
SRE       & 0.0727     & 0.0720     & 0.2023        &  & 0.0861     & 0.0867     & 0.3259        \\
OOS RiskE & 2.50E-04   & 1.89E-04   & 1.03E-03      &  & 2.48E-04   & 1.90E-04   & 1.49E-03      \\
OOS SRE   & 0.0382     & 0.0348     & 0.1339        &  & 0.0389     & 0.0349     & 0.1606        \\ \midrule
          & \multicolumn{3}{c}{Replace 4}           &  & \multicolumn{3}{c}{Replace 5}           \\
EcovE     & 4.94E-04   & 4.31E-04   & 1.26E-02      &  & 5.06E-04   & 5.02E-04   & 1.21E-02      \\
INVcovE   & 7617       & 5818       & 30268         &  & 7604       & 5899       & 575714        \\
WE        & 0.5187     & 0.4950     & 1.3918        &  & 0.5169     & 0.5008     & 2.8890        \\
RiskE     & 1.59E-04   & 1.32E-04   & 8.04E-04      &  & 1.58E-04   & 1.34E-04   & 9.65E-04      \\
SRE       & 0.0887     & 0.0899     & 0.3154        &  & 0.0856     & 0.0871     & 0.3545        \\
OOS RiskE & 2.48E-04   & 2.02E-04   & 2.01E-03      &  & 2.46E-04   & 2.07E-04   & 5.00E-03      \\
OOS SRE   & 0.0368     & 0.0332     & 0.1595        &  & 0.0377     & 0.0345     & 0.1507        \\ \midrule
          & \multicolumn{3}{c}{Replace 6}           &  &            &            &               \\
EcovE     & 5.70E-04   & 5.88E-04   & 1.22E-02      &  &            &            &               \\
INVcovE   & 7619       & 5965       & 62467         &  &            &            &               \\
WE        & 0.5162     & 0.5063     & 2.9828        &  &            &            &               \\
RiskE     & 1.57E-04   & 1.33E-04   & 9.35E-04      &  &            &            &               \\
SRE       & 0.0890     & 0.0911     & 0.3440        &  &            &            &               \\
OOS RiskE & 2.40E-04   & 2.12E-04   & 5.19E-03      &  &            &            &               \\
OOS SRE   & 0.0357     & 0.0330     & 0.1483        &  &            &            &       \\ \bottomrule       
\end{tabular}   
\end{table}

 We generate the $p_m \times T$ pseudo return data $\{r_t^{(m)}\}_{t=1}^{T}$ for sector $m$ based on the factor model defined in (\ref{yr(1)}).  For the $i^{th}$ common factor in the $m$ sector, we assume it follow a stationary AR(1) process, that is $f_{i,t}^{(m)} = 
 \mu_{i}^{(m)}+\alpha_i^{(m)} f_{i,t-1}^{(m)} + u_{i,t}^{(m)}$
 for $i=1,\ldots,K_m$ and $m = 1,\ldots, M$ with $f_{i,0}^{(m)}=0$. Here,  $u_{i,t}^{(m)}\sim N(0,1-\alpha_i^{(m)2})$, which indicates the covariance matrix of common factors in all sectors are identity matrix.  Similar setup for generating common factors can be found, e.g., in \cite{fan2013large} and \cite{fan2022time}. The number of common factors $K_m$ for each sector is set to 2. For factor loadings $b_i = \left( b_{i,1}^{(m)}, b_{i,2}^{(m)} \right)^\top$,  we draw the elements of loading matrix from normal distribution, i.e., $b_{i,j}^{(m)} \sim N(\mu_{b,j},\sigma_{b,j}^2), j = 1, 2$.   Finally, for the idiosyncratic error $e_{t}^{(m)} = (e_{1t}^{(m)},\ldots,e_{pt}^{(m)})^\top$, we generate the random errors from the multivariate normal distribution $ N(\bm{0}_{p_m},\Sigma_{e}^{(m)})$.  

 \begin{table}[htbp!]
 \caption{Simulation results for the scenario in which sectors have different number of assets, and the sample size is 125. EcovE is the error covariance estimation error, INVcovE is the inverse of return covariance matrix estimation error, WE is weight estimation error, RiskE is risk estimation error, SRE is the Sharpe ratio estimation error, OOS RiskE is the distance between out-of-sample estimated portfolio risk and that of the population portfolio, OOS SRE is the distance between out-of-sample estimated portfolio Sharpe ratio and that of the population portfolio. Replace 0-3 denote the number of replacement steps.
“Replace 0” corresponds to complete homogeneity, where all sectors share the same two basic factors. At each step, a block of four factors is replaced by newly generated ones. “Replace 3” corresponds to maximal heterogeneity, where all sector-specific factors are fully distinct.}
\label{Table Simu2}
\centering
\tabcolsep 0.056in
\begin{tabular}{cccccccc}
 \toprule      Measures   & Individual & Multi-task & Pooled-sector &  & Individual & Multi-task & Pooled-sector \\ \midrule
          & \multicolumn{3}{c}{Replace  0}          &  & \multicolumn{3}{c}{Replace 1}           \\
EcovE     & 1.19E-03   & 1.11E-03   & 1.14E-03      &  & 1.20E-03   & 1.12E-03   & 1.27E-02      \\
INVcovE   & 12316      & 10771      & 12262         &  & 12357      & 10823      & 55979         \\
WE        & 0.8296     & 0.8198     & 0.8185        &  & 0.8332     & 0.8195     & 32.3414       \\
RiskE     & 7.40E-05   & 6.56E-05   & 6.97E-05      &  & 7.61E-05   & 6.67E-05   & 3.12E-03      \\
SRE       & 0.0162     & 0.0150     & 0.0161        &  & 0.0312     & 0.0304     & 0.1891        \\
OOS RiskE & 4.02E-04   & 3.75E-04   & 3.99E-04      &  & 4.08E-04   & 3.79E-04   & 5.86E-02      \\
OOS SRE   & 0.0522     & 0.0508     & 0.0542        &  & 0.0564     & 0.0549     & 0.1001        \\ \midrule
          & \multicolumn{3}{c}{Replace 2}           &  & \multicolumn{3}{c}{Replace 3}           \\
EcovE     & 1.21E-03   & 1.13E-03   & 1.56E-02      &  & 1.22E-03   & 1.15E-03   & 1.54E-02      \\
INVcovE   & 12339      & 10829      & 192130        &  & 12345      & 10962      & 249821        \\
WE        & 0.8307     & 0.8186     & 18.8892       &  & 0.8313     & 0.8187     & 13.2574       \\
RiskE     & 7.51E-05   & 6.55E-05   & 2.88E-03      &  & 7.55E-05   & 6.51E-05   & 2.60E-03      \\
SRE       & 0.0480     & 0.0474     & 0.4015        &  & 0.0527     & 0.0545     & 0.3933        \\
OOS RiskE & 4.07E-04   & 3.79E-04   & 3.64E-02      &  & 4.10E-04   & 3.81E-04   & 2.48E-02      \\
OOS SRE   & 0.0592     & 0.0577     & 0.1226        &  & 0.0517     & 0.0508     & 0.1208   \\ \bottomrule    
\end{tabular}
\end{table}  
	
To better mimic real-world investment decision environment, all population parameter values in the above setup are calibrated using real financial data.  Specifically, we collect daily return data for the constituent stocks of the S\&P 500 index over the period from January 2006 to December 2009, obtained from the CRSP database. We then select the largest 250 stocks based on market capitalization and apply the classical PCA procedure (with the number of factor 2) to estimate the factor loadings $\hat{B}$ and common factors $\hat{F}$. Based on these estimates, we further obtain the sparse error covariance matrix estimation $\hat{\Sigma}_e$  using the adaptive sparse thresholding method of \cite{cai2011adaptive} with a thresholding parameter of 0.5. Then, we set $\mu_{b,1}^{(m)}  = 0.0195, \mu_{b,2}^{(m)} = -0.001, \sigma_{b,1}^{(m)}= 0.0081, \sigma_{b,2}^{(m)}= 0.0079$ according to the mean and variance of $\hat{B}$.  Since we aim to examine how different degrees of heterogeneity (or homogeneity) of the factor spaces across sectors affect the performance of the newly proposed method, we construct  data-generating processes in which the level of heterogeneity is gradually increased from the case of complete homogeneity. In the complete homogeneity case, all sectors share the same common factors, which we refer to as the basic factors. We then introduce heterogeneity by progressively replacing the factors in each sector with newly generated ones. For ease of exposition, we refer to these newly generated factors as replaced factors. As the number of replaced factors increases, the degree of heterogeneity across sectors becomes stronger.
For the autoregressive coefficients $\alpha_i^{(m)}$ of the two basic  factors,  we set them equal to the AR(1) least squares estimates obtained from the estimated factors $\hat{F}_t$, which are $-0.0838$ and $-0.0206$, respectively. For the replaced factors, the coefficients are randomly and uniformly drawn from $[-0.1,0.1]$. 
For the intercept term $\mu_i^{(m)}$, we draw its values independently from a uniform distribution on $[0,1]$. Lastly, for $\Sigma_{e}^{(m)}$ across different sectors, we set them to the sparse error covariance matrix estimators $\hat{\Sigma}_e^{(m)}$, computed using $p$ randomly selected stocks over any one-year period.

 We evaluate the performance of the proposed multi-task portfolios (in the column ``Multi-task'' of Tables \ref{Table Simu1} and \ref{Table Simu2}) under varying degrees of homogeneity. Specifically, we begin with the case of complete homogeneity, in which we consider $M = 16$ sectors that share the same two basic common factors.  For ease of exposition, we denote the two basic factors as $F_1$ and $F_2$, and thus $F^{(m)}_i = F_i$ for $i=1,2$ and $m = 1,\ldots,16$ in the baseline setting.
 Then we gradually increase the heterogeneity among sectors: we sequentially replace a block of  five factors in $\{F^{(2)}_1,F^{(3)}_1,\ldots,F^{(16)}_1,F^{(2)}_2,F^{(3)}_2,\ldots, F^{(16)}_2 \}$ with newly generated replaced factors, one block at a time. For example, in the first replacement step, the first five factors $F^{(2)}_1,F^{(3)}_1,\ldots, F^{(6)}_1$, which are all equal to $F_1$ in the baseline setting, are replaced  by new factors $F^{(2)*}_1, F^{(3)*}_1,\ldots, F^{(6)*}_1$. 
After six such replacements (i.e., $6\times 5=30$ factors are replaced), all 16 sectors are completely distinct from each other. Consequently, the data-generating process transitions gradually from complete homogeneity to maximal heterogeneity. The number of assets in each sector is set to 50, and the sample size is 250. In addition, we consider a high-dimensional setting in which the number of assets exceeds the sample size in some sectors. In this case, we set $M= 7$, and the numbers of assets across the seven sectors are 50, 75, 100, 125, 150, 175, and 200, with a sample size of  $T =125$. In each replacement step, we replace four factor.

For performance evaluation, we consider the following measures.
\begin{itemize}
    \item[(1)] \textbf{E}rror \textbf{cov}ariance matrix estimation \textbf{E}rror (EcovE): the distance between  the estimated error covariance matrix  and  the corresponding population counterpart measured by spectral norm, i.e., $\text{EcovE}^{(m)} = \left\|\hat{\Sigma}_e^{(m)} - \Sigma_e^{(m)} \right\|$.
    \item[(2)] \textbf{INV}erse of  \textbf{cov}ariance matrix estimation \textbf{E}rror (INVcovE): the distance between  the inverse of estimated return covariance matrix  and  the corresponding population counterpart measured by spectral norm, i.e., $\text{INVcovE}^{(m)} = \left\|\hat{\Omega}_r^{(m)} - \Omega_r^{(m)} \right\|$.
    \item[(3)] \textbf{W}eight estimation \textbf{E}rror (WE):  the $\ell_1$ norm of the difference between the estimated portfolio weights and oracle version $\text{WE}^{(m)} = \mathop{\Vert\widehat{w}^{(m)}-w^{*(m)}\Vert}_1 $.
    \item[(4)] \textbf{Risk} estimation \textbf{E}rror (RiskE): the absolute difference between the portfolio risk estimation and the corresponding true level, i.e., $\text{RiskE}^{(m)} = \left|\sqrt{\hat{R}_{min}^{(m)}} - \sqrt{R_{min}^{(m)}}\right|$.
    \item[(5)] \textbf{S}harpe \textbf{R}atio estimation \textbf{E}rror (SRE): the absolute difference between the  portfolio Sharpe ratio estimation and the corresponding true level, i.e., $\text{SRE}^{(m)} = \left|\widehat{SR}^{(m)} - SR^{(m)}\right|$.

    \item[(6)] \textbf{O}ut-\textbf{O}f-\textbf{S}ample \textbf{R}isk \textbf{E}rror (OOS RiskE): the absolute difference between the out-of-sample risk of estimated portfolio and the population portfolio, i.e., $\text{OOS RiskE}^{(m)} = \left|\sqrt{\hat{w}^{(m)\top}\Sigma_r^{(m)}\hat{w}^{(m)}} - \sqrt{R_{min}^{(m)}}\right|.$
    
    \item[(7)] \textbf{O}ut-\textbf{O}f-\textbf{S}ample \textbf{S}harpe \textbf{R}atio  \textbf{E}rror (OOS SRE): the absolute difference between the out-of-sample SR of estimated portfolio and the population portfolio, i.e., 
    $\text{OOS SRE}^{(m)} = \left|\hat{w}^{(m)\top}\mu_r^{(m)}/\sqrt{\hat{w}^{(m)\top}\Sigma_r^{(m)}\hat{w}^{(m)}} - SR^{(m)}\right|$.
\end{itemize} 
In the above definitions, $\hat{A}$ denotes the estimator of $A$, obtained from different strategies.
We report overall performance, where all results are averaged across the 16 sectors.

For benchmarks, we compare   with the individual strategy and the pooled-sector strategy.  The \textbf{individual strategy} constructs a minimum-variance portfolio for each sector based on the POET covariance estimator obtained using only the data from that sector.
The \textbf{pooled-sector strategy} estimates the common factors and factor loadings using pooled data across all sectors; the covariance matrix for each sector is then constructed using the estimated factor component together with the sector-specific error covariance matrix, where the latter is obtained by applying adaptive sparse thresholding method to the estimated residuals within each sector; the corresponding  minimum-variance portfolio for each sector is then formed by using the sector-specific covariance estimation. The regularization parameter $\lambda$ for our new strategy  is selected via the cross-validation procedure described in Subsection \ref{sec: choice of regu}, and the sparse threshold parameter $C_\tau$ is selected by data-driven method defined in \eqref{eq:Ctau select}.\footnote{For the other two benchmarks that also involve sparse error covariance matrix estimation, we use the same value of $C_\tau$ as in our method.} The number of common factors in each sector is assumed to be known, and the number of simulation replications is set to 1000.

\subsection{Simulation Results}

Table \ref{Table Simu1} reports the simulation results  under the scenario where all sectors have the same number of assets. From the table, it can be seen that the proposed multi-task learning-based estimation procedure achieves the best performance in terms of both matrix-related and portfolio-related measures in most cases. For example, under the partial homogeneity setting “Replace 2”, the estimation errors for the error covariance matrix, the inverse of return covariance matrix, portfolio weights, portfolio risk, and the Sharpe ratio are 3.31e-04, 5581, 0.4803, 1.26e-04 and 0.072, respectively, all of which are the lowest. Meanwhile, the proposed portfolio strategy also achieves the lowest out-of-sample risk error (1.89e-04) and Sharpe ratio error (0.0348). Furthermore, under the complete homogeneity setting, the pooled-sector strategy outperforms the individual strategy. However, its performance deteriorates sharply as heterogeneity is introduced across sectors. From the complete homogeneity setting to “Replace 1”, the out-of-sample risk error and Sharpe ratio error of the pooled-sector portfolio increase from 2.10E-04 to 5.55E-04, and from  0.0318 to 
0.0717, respectively.
Under the complete heterogeneity setting, the performance of the proposed strategy is comparable to that of the individual strategy, although it may exhibit slight disadvantages with respect to certain measures, it outperforms in others. More importantly, it is clear that under transition settings (i.e., partial homogeneity), which are likely the most common scenarios in practice, the proposed multi-task learning-based strategy consistently outperforms the individual strategy and pooled-sector strategy in most cases.

Table \ref{Table Simu2} reports the simulation results for the case where the number of assets in some sectors exceeds the sample size. From Table \ref{Table Simu2}, similar conclusions can be drawn as in the previous setting, and we therefore omit a similar discussion. It is worth noting that, even under complete homogeneity, the pooled-sector strategy does not beat the individual strategy. This can be attributed to the limited sample size, which plays a leading role in determining estimation accuracy. As shown in \cite{fan2013large} and our theoretical results, the convergence rate is driven by the term   $1/\sqrt{p}+\sqrt{\log p/T}$. 
Consequently, for both individual strategy and pooled-sector strategy (as well as our method), they are constrained by the sample size under the second setup, so that increasing the number of assets (the benefit of pooling data together) yields only limited improvement in estimation accuracy. In contrast, in the previous setting, the number of assets (50) is smaller than the sample size (250),  the performance of individual strategy is mainly constrained by the cross-sectional dimension, which can be alleviated by pooling data across sectors.

Finally, one can observe that the advantage of the proposed strategy becomes more pronounced as the degree of homogeneity increases. For example, moving from complete heterogeneity to complete homogeneity, the gap in the inverse covariance matrix estimation error (INVCovE) between the proposed method and the individual strategy increases, with values given by 1654, 1706, 1799, 2030, 2067, 2268, and 2859. This pattern indicates that the degree of superiority of the proposed strategy (over individual strategy) depends on the level of homogeneity, with greater gains achieved in more homogeneous settings.

\section{Empirical Analysis}
\label{sec: empirical}

In this section, we use real data to demonstrate the implementation of the proposed multi-task strategy and evaluate its practical performance. Specifically, the empirical analysis is based on daily return data of individual stocks from the Russell 3000 index, obtained from the CRSP database. It is common in the real investment to have themed funds focusing on some specific sector. In this study, we adopt the Fama-French industrial sectors, as detailed below. The evaluation is conducted under a rolling window framework. At each decision point, investors construct the portfolio using the historical data available up to that time. The portfolio is then held for the subsequent holding period. Afterward, the estimation window rolls forward to the next decision point, where the portfolio is re-estimated using the updated information set. This procedure is repeated throughout the entire sample period. We evaluate the performance of various portfolios using three out-of-sample performance measures: the out-of-sample Sharpe ratio, out-of-sample risk, and out-of-sample cumulative excess return. These measures are defined  as follows respectively:
\begin{align*}
\text{CR} &= \sum_{t=1}^{\mathbb{T}} r_{t}^{p},  \quad 
\text{Risk} = \sqrt{\frac{1}{\mathbb{T}-1}\sum_{t=1}^\mathbb{T}\left(r_t^p - \frac{\text{CR}}{\mathbb{T}}\right)^2}, \quad  
\text{SR} = \frac{\text{CR}}{\mathbb{T} \times \text{Risk}}
\end{align*}
where $\mathbb{T}$ is the number of out-of-sample observations and $r_t^p$ is out-of-sample portfolio excess return.  We exclude stocks from the portfolio that lack a complete return history for the specified window length at the initial decision point. However, stocks that exhibit missing values in some but not all subsequent windows are retained to avoid look-ahead bias in the rolling window procedure \citep{ao2019approaching, barroso2022lest}. The risk-free rate is obtained from the Fama-French data library.

\begin{table}[htbp]
\centering
\caption{The out-of-sample overall performance of the newly proposed portfolio and various benchmarks.  We allocate stocks in the Russell 3000 index to the sectors \textit{NoDur}, \textit{Durbl}, \textit{Manuf}, \textit{Enrgy}, \textit{HiTec}, \textit{Telcm}, \textit{Shops}, \textit{Hlth}, and \textit{Utils} according to the Fama-French 10 industry classification. Daily excess return data is applied under a rolling window scheme with a sample size of 252. Portfolios are rebalanced monthly (21 days), and the out-of-sample period is from 04/01/2010 to 31/12/2024. CM is the out-of-sample cumulative excess return, Risk is the out-of-sample standard deviation of portfolio return, SR is the out-of-sample portfolio Sharpe ratio. }
\label{Table 252 basic}
\tabcolsep 0.056in
\begin{tabular}{ccccccc}
\toprule
Scenario & Measures & Multi-task     & Individual     & Pooled-sector    & Global pooling    & Equal weights    \\ \midrule
(1) &      & \multicolumn{5}{c}{\textit{Manuf} and   \textit{Enrgy}}                                 \\
    & CM   & 1.5792           & 1.4471           & -10.7063         & 0.456           & 2.1744         \\
    & Risk & 0.00700           & 0.00726           & 0.20126           & 0.01632         & 0.01582        \\
    & SR   & 0.0601           & 0.0531           & -0.0142          & 0.0074          & 0.0366         \\ \midrule
(1) &      & \multicolumn{5}{c}{\textit{NoDur}, \textit{Durbl}, and \textit{Shops}}                  \\
    & CM   & 1.808            & 1.792            & 1.719            & 1.575           & 2.2292         \\
    & Risk & 0.00679          & 0.00704          & 0.00690          & 0.00643         & 0.01335        \\
    & SR   & 0.0708           & 0.0678           & 0.0663           & 0.0652          & 0.0444         \\ \midrule
(1) &      & \multicolumn{5}{c}{\textit{HiTec}, \textit{Telcm}, \textit{Hlth}, and   \textit{Utils}} \\
    & CM   & 1.4056           & 1.423            & 1.498            & 1.781           & 2.2274         \\
    & Risk & 0.00670          & 0.00694          & 0.00680          & 0.01731         & 0.01188        \\
    & SR   & 0.0559           & 0.0545           & 0.0586           & 0.0274          & 0.0499         \\ \midrule
(2) &      & \multicolumn{5}{c}{5 subsectors in \textit{Manuf} }                                     \\
    & CM   & 1.898            & 1.889            & 1.893            & 1.782           & 2.3022         \\
    & Risk & 0.00780          & 0.00806          & 0.00801          & 0.00751         & 0.01425        \\
    & SR   & 0.0648           & 0.0624           & 0.0629           & 0.0631          & 0.0430         \\ \midrule
(2) &      & \multicolumn{5}{c}{5 subsectors in \textit{HiTec}}                                      \\
    & CM   & 2.0573           & 1.7930           & 1.9570           & 1.7636          & 2.5393         \\
    & Risk & 0.00920          & 0.00928          & 0.00929          & 0.00792         & 0.01444        \\
    & SR   & 0.0595           & 0.0514           & 0.0561           & 0.0593          & 0.0468      \\ \bottomrule    
\end{tabular}
\end{table}

We compare the performance of the proposed multi-task learning-based portfolio with those of the individual strategy, the pooled-sector strategy (both defined in the simulation section), and the global pooling strategy. The \textbf{global pooling strategy} estimates the common factors and factor loadings using pooled data across all sectors, and further estimates a global sparse error covariance matrix using the entire set of residuals. Based on this covariance estimator, a global minimum-variance portfolio is constructed using all available stocks.
In addition, we also report the results of equal weighting strategy \citep{demiguel2009generalized, yuan2024naive}, which is a benchmark in high-dimensional portfolio allocation.
We  evaluate the overall performance over all sectors  under an equal allocation scheme. Specifically, for the multi-task strategy, the individual strategy, the pooled-sector strategy and equal weighting strategy, we allocate equal initial capital to each sector, after which each sector portfolio evolves independently. Accordingly, the overall portfolio performance is obtained by aggregating sector-level net asset values (NAVs), where each sector-level NAV is constructed by cumulatively compounding its daily returns. The resulting aggregate NAV is then used to compute the overall portfolio returns, and thus compute the corresponding out-of-sample performance.
The regularization parameter $\lambda$  is selected by the cross-validation (\ref{cv for lambda}) and is updated while rebalancing portfolio. The number of common factors and the values of  thresholding parameter $C_\tau$ for all strategies are determined  by equations \eqref{eq: factor number esti} and \eqref{eq:Ctau select} at the first decision node, respectively.  For multiple sectors, we group individual stocks according to the Fama-French 10 industry classification, which has been widely used in the financial literature and is known to better capture stock return co-movements. Specifically, the 10 Fama-French industry classifications are \textit{NoDur} (consumer Nondurables including food, tobacco, textiles, apparel, leather, and toys), \textit{Durbl} (consumer Durables including cars, TVs, furniture, and household Appliances), \textit{Manuf} (Manufacturing including machinery, trucks, planes, chemicals, off furn, paper, com printing), \textit{Enrgy} (oil, gas, and coal extraction and Products), \textit{HiTec} (Business Equipment including computers, software, and electronic equipment), \textit{Telcm} (telephone and television transmission
), \textit{Shops} (wholesale, retail, and some services such as laundries, repair shops), \textit{Hlth} (healthcare, medical equipment, and drugs), \textit{Utils} (Utilities) and \textit{Other}.\footnote{The detailed SIC code ranges corresponding to each industry classification follow the definitions provided in the Fama-French industry classification scheme. see \url{https://mba.tuck.dartmouth.edu/pages/faculty/ken.french/Data_Library/det_10_ind_port.html}.}

\begin{table}[htbp!]
\centering
\caption{The out-of-sample overall performance of the newly proposed portfolio and various benchmarks. We allocate stocks in the Russell 3000 index to the sectors \textit{NoDur}, \textit{Durbl}, \textit{Manuf}, \textit{Enrgy}, \textit{HiTec}, \textit{Telcm}, \textit{Shops}, \textit{Hlth}, and \textit{Utils} according to the Fama-French 10 industry classification. Daily excess return data is applied under a rolling window scheme with a sample size of 504. Portfolios are rebalanced monthly (21 days), and the out-of-sample period is from 04/01/2010 to 31/12/2024. CM is the out-of-sample cumulative excess return, Risk is the out-of-sample standard deviation of portfolio return, SR is the out-of-sample portfolio Sharpe ratio. }
\label{Table scenario 504}
\tabcolsep 0.056in
\begin{tabular}{ccccccc}
\toprule 
Scenario & Measures & Multi-task     & Individual     & Pooled-sector    & Global pooling    & Equal weights    \\ \midrule
(1)      &          & \multicolumn{5}{c}{\textit{Manuf} and \textit{Enrgy}}                                   \\
   & CM   & 1.5795           & 1.5194           & 1.4970          & 1.7970          & 2.1793          \\
    & Risk & 0.00719          & 0.00724          & 0.00748         & 0.01243         & 0.01578         \\
    & SR   & 0.0584           & 0.0558           & 0.0532          & 0.0385          & 0.0367          \\ \midrule
(1) &      & \multicolumn{5}{c}{\textit{NoDur}, \textit{Durbl}, and \textit{Shops}}                  \\
    & CM   & 1.7220           & 1.7427           & 1.7206          & 1.5729          & 2.2399          \\
    & Risk & 0.00723          & 0.00739          & 0.00734         & 0.00668         & 0.01330         \\
    & SR   & 0.0634           & 0.0627           & 0.0624          & 0.0627          & 0.0448          \\ \midrule
(1) &      & \multicolumn{5}{c}{\textit{HiTec}, \textit{Telcm}, \textit{Hlth}, and   \textit{Utils}} \\
    & CM   & 1.3516           & 1.3287           & 1.3509          & 3.1066          & 2.2295          \\
    & Risk & 0.00689          & 0.00727          & 0.00708         & 0.02502         & 0.01188         \\
    & SR   & 0.0522           & 0.0486           & 0.0508          & 0.0330          & 0.0499          \\ \midrule
(2) &      & \multicolumn{5}{c}{5 subsectors in \textit{Manuf} }                                     \\
    & CM   & 1.9866           & 1.9274           & 1.9294          & 1.8213          & 2.3079          \\
    & Risk & 0.00794          & 0.00819          & 0.00810         & 0.00709         & 0.01425         \\
    & SR   & 0.0666           & 0.0626           & 0.0634          & 0.0683          & 0.0431          \\ \midrule
(2) &      & \multicolumn{5}{c}{5 subsectors in \textit{HiTec}}                                      \\
    & CM   & 2.0551           & 1.7712           & 1.9419          & 1.5730          & 2.5399          \\
    & Risk & 0.00960          & 0.00946          & 0.00960         & 0.00776         & 0.01444         \\
    & SR   & 0.0569           & 0.0498           & 0.0538          & 0.0539          & 0.0468         \\ \bottomrule  
\end{tabular}
\end{table}

Based on this classification, we consider two practical scenarios in which investors jointly allocate portfolios across multiple sectors and focus on the overall performance. (1) In the first scenario, investors make portfolio decisions jointly for \textit{NoDur}, \textit{Durbl}, and \textit{Shops}, which are all treated as consumer-related sectors; for \textit{Manuf} and \textit{Enrgy}, which represent production-oriented sectors; and for the remaining sectors, \textit{HiTec}, \textit{Telcm}, \textit{Hlth}, and \textit{Utils}, which are characterized by their provision of technology and essential infrastructure services. 
Under the multi-task framework, we expect to utilize the similarity across various sectors  to improve overall performance of multiple portfolio decisions. Accordingly, economically related sectors are grouped together and treated as a unified investment problem. In practice, investors can freely choose the sectors in which they invest. Our objective here is not to identify the optimal sector combination in terms of the Sharpe ratio, but rather to demonstrate that the proposed method achieves strong overall performance when some a priori given sectors (e.g., the 10 Fama-French industrial sectors in our example) share homogeneous information. (2) In the second scenario, we further divide the \textit{Manuf} and \textit{HiTec} sectors into five subsectors each. For the \textit{Manuf} sector, the classification reflects broad similarities in production activities, including light manufacturing, chemical and processing industries, industrial machinery, electrical equipment, and transportation-related manufacturing.\footnote{The SIC codes for each subgroup in \textit{Manuf} are as follows: subgroup 1 includes SIC 2520–2589 and 2600–2699; subgroup 2 includes SIC 2750–2769, 2800–2829, and 2840–2899; subgroup 3 includes SIC 3000–3099 and 3200–3569; subgroup 4 includes SIC 3580–3621 and 3623–3629; and subgroup 5 includes SIC 3700–3709, 3712–3713, 3715, 3717–3749, 3752–3791, 3793–3799, and 3860–3899.} This grouping introduces a  degree of within-sector granularity.  For the \textit{HiTec} sector, we classify stocks into five categories corresponding to major segments of the technology industry: computer hardware, electronic equipment, precision instruments, software and IT services, and research and development services.\footnote{The SIC codes for each subgroup in \textit{HiTec} are defined as follows: subgroup 1 corresponds to SIC 3570–3579; subgroup 2 corresponds to SIC 3622, 3660–3692, and 3694–3699; subgroup 3 corresponds to SIC 3810–3839; subgroup 4 corresponds to SIC 7370–7379; and subgroup 5 corresponds to SIC 7391 and 8730–8734.} 
 The number of assets varies across sectors, the \textit{Manuf} sector has the largest number of assets 400, while the \textit{Durbl} sector has the smallest number 52 at the initial decision point. And we consider window sizes of 252 (one year of historical data) and 504 (two years). In the former case, the number of assets in some sectors (e.g., \textit{Manuf} and \textit{HiTec}) exceeds the sample size, whereas in the latter case the sample size is larger than the number of assets in each sector. We rebalance portfolios monthly (every 21 days), and the out-of-sample period spans from 01/2010 to 12/2024, covering a total of 15 years and including the COVID-19 outbreak.

Tables \ref{Table 252 basic}-\ref{Table scenario 504} report the empirical results of all considered strategies.  First of all, it  can be observed that the newly proposed multi-task portfolio achieves the highest Sharpe ratio in most cases, compared to the the portfolios constructed from the individual strategy and the two pooled strategy, indicating that the newly proposed strategy can deliver ideal overall performance. 
Second, in terms of portfolio risk, the proposed strategy outperforms both the individual strategy and the pooled-sector strategy in most cases. We note that the global pooling strategy attains the lowest risk in some cases, which is expected, particularly in Scenario (2) where the degree of homogeneity is high, as it corresponds to the global minimum variance portfolio. However, it also exhibits substantially higher risk in Scenario (1), where the sectors display a greater degree of heterogeneity. For example, in Scenario (1) with investment in the \textit{Manuf} and \textit{Energy} sectors, its risk is 0.01632, which is much larger than that of our method, 0.007. Third, the equal weighting strategy is also dominated by our new strategy in terms of risk and SR among all  cases. Fourth, both pooled strategies may perform poorly in the first scenario but generally perform well in the second scenario. For example, when practitioners invest in stocks in \textit{Manuf} and \textit{Energy}, the pooled-sector strategy and the global pooling strategy obtain Sharpe ratios of -0.0142 and 0.0074, respectively, which are both significantly lower than that of our portfolio (0.0601); when practitioners invest in the five subsectors in \textit{Manuf}, the two pooled strategies achieve Sharpe ratios of 0.0629 and 0.0631, respectively, which are comparable to our method. It suggests that sectoral heterogeneity may substantially deteriorate the performance of the pooled strategy.


\begin{table}[htbp!]
\centering
\caption{Empirical portfolio performance for the \textit{Servs}, \textit{BusEq}, and \textit{FabPr} sectors under the Fama-French 30 industry classification. The \textit{Servs} sector includes firms in personal and business services, the \textit{BusEq} sector consists of firms in business equipment, and the \textit{FabPr} sector comprises firms in fabricated products and machinery.  For each of three sector, we incorporate highly correlated Fama-French 30 industry sectors (with a minimum correlation of 0.95, determined by the linear correlation between equal-weighted portfolios of two sectors at the initial decision node) to enhance homogeneity.}
\label{Table corre 95 results 2}
\begin{tabular}{ccccc}
\toprule 
Measures     & Multi-task           & Individual           & Pooled-sector        & Equal weights        \\ \midrule
     & \multicolumn{4}{c}{\textit{Servs}}                                                                 \\ 
CM   & 1.951                & 1.723                & 1.868                & 2.559                \\
Risk & 0.00818              & 0.00832              & 0.00814              & 0.01527              \\
SR   & 0.0635               & 0.0551               & 0.0611               & 0.0446               \\ \midrule
     & \multicolumn{4}{c}{\textit{BusEq}}                                                                 \\ 
CM   & 2.197                & 2.156                & 2.208                & 2.634                \\
Risk & 0.00945              & 0.00972              & 0.00966              & 0.01321              \\
SR   & 0.0619               & 0.0590               & 0.0608               & 0.0531               \\ \midrule
& \multicolumn{4}{c}{\textit{FabPr}}                 \\ 
CM   & 2.071                & 2.229                & 2.003                & 2.544                \\
Risk & 0.00799              & 0.00904              & 0.00787              & 0.01304              \\
SR   & 0.0690               & 0.0656               & 0.0678               & 0.0519     \\ \bottomrule         
\end{tabular}
\end{table}

In practice, investors may also be interested in portfolio allocation within a specific sector. Therefore, we next focus on one sector and evaluate the potential performance enhancement from incorporating information from other related sectors. Specifically, we conduct multi-task learning for a sector of interest by selecting and incorporating other sectors that are quite highly correlated with it, so that the homogeneous information can help improve portfolio performance for the sector of interest. 
For simplicity, we measure the correlation between sectors using the linear correlation of the excess returns from their equal weighting portfolios at the initial decision point. The threshold for high correlation is set to 0.95. In the supplementary material, we also report the  results using the distance between the projection matrices spanned by the common factors of two sectors to characterize homogeneous information, which leads to similar results. 
As the 10 classification provides too few  groups for this analysis, we further adopt the Fama-French 30 industry classification and group the stocks in the Russell 3000 index into 30 sectors.\footnote{The sectors with fewer than 20 stocks at the first portfolio construction are removed to ensure that each sector contains enough stocks throughout the rolling window procedure. As such, we finally obtain 25 sectors. The sectors of \textit{Beer}, \textit{Smoke}, \textit{Textiles}, \textit{Carry}, and \textit{Coal} are excluded.}  In this new scenario, we focus on the portfolio performance within the target sector and investigate the effect of positive transfer,\footnote{We say that positive transfer exists for a target task if the out-of-sample performance based on equation \eqref{PCA loss Multi} improves upon that of training the target task alone.} a key aspect also emphasized in multi-task learning research \citep{wu2020understanding}.

Specifically, we consider three sectors \textit{FabPr} (Fabricated Products and Machinery), \textit{Servs} (Personal and Business Services), and \textit{BusEq} (Business Equipment) in the main text while the analogous study can be applied to other sectors.  
The corresponding empirical results for each sector are tabulated in Table \ref{Table corre 95 results 2}. Since the global pooling strategy is applied to all stocks and is not available at the sector level, we ignore it. From Table \ref{Table corre 95 results 2}, it is evident that the multi-task portfolio in each sector obtains a higher portfolio SR and lower risk than the corresponding portfolios from the individual strategy, which indicates the existence of positive transfer. Furthermore, the new portfolio also achieves better performance than the pooled-sector strategy, even when the sectors considered in this scenario are highly correlated. We note that the analysis based on pooled data is largely affected by the heterogeneity across sectors, as indicated by our simulation results, even when such heterogeneity is small.
Practitioners focusing on the target sector can leverage the proposed multi-task learning approach to achieve improved returns relative to risk.

\section{Conclusion}
\label{sec: con}
This paper proposes a data-adaptive, multi-sector factor modeling for asset returns. The key idea is to exploit potential homogeneity across sectors, which are often interconnected, to improve the overall portfolio performance.  We establish theoretical guarantees for the proposed approach. Numerical results demonstrate its favorable finite-sample performance and its practical applicability across a range of empirical settings.

\bibliographystyle{chicago}
\bibliography{reference}

\end{sloppypar}

\end{document}